\documentclass[10pt, twocolumn]{IEEEtran}
\usepackage{url}
\usepackage{graphicx}
\usepackage{algorithm}
\usepackage[noend]{algpseudocode}
\usepackage{etex} 
\usepackage{tikz}
\usetikzlibrary{shapes,arrows}
\usepackage{pstricks}
\usepackage{pst-node,pst-blur}
\usetikzlibrary{arrows}
\usepackage{amsfonts}
\usepackage{tabularx}
\usepackage{subfigure}
\usepackage{amssymb}
\usepackage{booktabs}
\usepackage{multirow}
\usepackage{epsfig}
\usepackage{epstopdf}

\usepackage[noadjust]{cite}
\newtheorem{theorem}{Theorem}
\newtheorem{remark}{Remark}

\newtheorem{lemma}{Lemma}

\newtheorem{corollary}{Corollary}

\usepackage[cmex10]{amsmath}
\usepackage[cmex10]{mathtools}
\DeclareMathOperator*{\argmin}{arg\,min}
\DeclarePairedDelimiter{\ceil}{\lceil}{\rceil}

\allowdisplaybreaks  

\makeatletter 
\def\@eqnnum{{\normalsize \normalcolor (\theequation)}} 
\makeatother
\begin{document}
\title{Energy Sustainable IoT with Individual QoS Constraints Through MISO SWIPT Multicasting}
\author{Deepak Mishra,~\IEEEmembership{Member,~IEEE}, George~C.~Alexandropoulos,~\IEEEmembership{Senior~Member,~IEEE}, \\and Swades De,~\IEEEmembership{Senior~Member,~IEEE}
\thanks{D. Mishra is with the Division of Communication Systems, Department of Electrical Engineering (ISY), Link\"oping University, Link\"oping 58183, Sweden (e-mail: deepak.mishra@liu.se).}
\thanks{G. C. Alexandropoulos is with the Mathematical and Algorithmic Sciences Lab, Paris Research Center, Huawei Technologies France SASU, 92100 Boulogne-Billancourt, France (e-mail: george.alexandropoulos@huawei.com). 
}
\thanks{S. De is with the Department of Electrical Engineering and Bharti School of Telecommunication, Indian Institute of Technology Delhi, New Delhi 110016, India (e-mail: swadesd@ee.iitd.ac.in).}
\thanks{{A preliminary version~\cite{ICC18} of this work has been accepted for presentation at IEEE ICC, Kansas City, USA, May 2018.}}}
 
\maketitle

\begin{abstract}
Enabling technologies for energy sustainable Internet of Things (IoT) are of paramount importance since the proliferation of high data communication demands of low power network devices. In this paper, we consider a Multiple Input Single Output (MISO) multicasting IoT system comprising of a multiantenna Transmitter (TX) simultaneously transferring information and power to low power and data hungry IoT Receivers (RXs). Each IoT device is assumed to be equipped with Power Splitting (PS) hardware that enables Energy Harvesting (EH) and imposes an individual Quality of Service (QoS) constraint to the downlink communication. We study the joint design of TX precoding and IoT PS ratios for the considered MISO Simultaneous Wireless Information and Power Transfer (SWIPT)
multicasting IoT system with the objective of maximizing the minimum harvested energy among IoT, while satisfying their individual QoS requirements. In our novel EH fairness maximization formulation, we adopt a generic Radio Frequency (RF) EH model capturing practical rectification operation, and resulting in a nonconvex optimization problem. For this problem, we first present an equivalent semi-definite relaxation formulation and then prove it possesses unique global optimality. We also derive tight upper and lower bounds on the globally optimal solution that are exploited in obtaining low complexity algorithmic implementations for the targeted joint design. Analytical expressions for the optimal TX beamforming directions, power allocation, and IoT PS ratios are also presented. Our representative numerical results including comparisons with benchmark designs corroborate the usefulness of proposed framework and provide useful insights on the interplay of critical system parameters.
\end{abstract} 
\begin{IEEEkeywords}
Energy harvesting, internet of things, multicasting, multiple antennas, optimization, power allocation, power splitting, simultaneous wireless information and power transfer.
\end{IEEEkeywords}
\IEEEpeerreviewmaketitle 
  

\section{Introduction and Background}\label{sec:introduction} 
\color{black}
Wireless Energy Harvesting (EH) has been recently considered as a key technological concept for the energy sustainability of the Internet of Things (IoT) \cite{IoT-Mag,ComMag}. An efficient technology belonging into this concept is the Simultaneous Wireless Information and Power Transfer (SWIPT) that targets at realizing perpetual operation of low power and data hungry network nodes \cite{IoT-P2,IoT-P3}. However, to achieve the goal of energy sustainable IoT via SWIPT, the fundamental bottlenecks of the practically available EH circuits need to be effectively handled. {Among these bottlenecks are the low rectification efficiency in Radio Frequency (RF) to Direct Current (DC) conversion and the relatively low receive energy sensitivity~\cite{ComMag}; the latter depends strongly on the distance between the Transmitter (TX) power node and a Receiver (RX) EH node.} Recent advances in multiantenna signal processing techniques for SWIPT~\cite{MIMO_SWIPT, MU-MISO-TBF-PS,MUTxMIMO,EBF2,CL17_MISO_SWIPT,MU_MISO_Access,MU_MISO_Access2,TX-PS,Close,MISO-Sep,EBF1,Rank1WCL,EH-BF} have revealed that the effective exploitation of the spatial dimension has the potential to overcome EH bottlenecks both in point-to-point systems and in multipoint communication like IoT. {Thus, SWIPT from a multiantenna TX has increased potential in providing continuous replenishment of the drained energy. However,  novel low complexity designs are needed to optimize the harvested power fairness among IoT nodes, while meeting their Quality of Service (QoS) demands~\cite{R2-IoTJ,R3-IoTJ}.}

\subsection{State-of-the-Art}\label{sec:RW}
In the seminal work~\cite{MIMO_SWIPT} focusing on the efficiency optimization of point-to-point multiantenna SWIPT systems, the trade off between achievable rate and received power for EH (also known as rate-energy trade off) was investigated for practical RX architectures. Power Splitting (PS), Time Switching (TS), and Antenna Switching (AS) architectures were proposed with the latter two being special cases of the former. Further, spatial switching architecture was recently investigated for QoS-aware harvested power maximization in MIMO SWIPT system~\cite{WCL18}. Based on these architectures, a lot of recent developments have lately appeared intending at enhancing the rate-energy performance of multiuser Multiple Input Single Output (MISO) SWIPT systems~\cite{MU-MISO-TBF-PS,MUTxMIMO,EBF2,CL17_MISO_SWIPT,MU_MISO_Access,MU_MISO_Access2,TX-PS,Close,MISO-Sep,EBF1,Rank1WCL,EH-BF, K-Tier}. These works mainly target at the optimization of TX precoding and PS operation, and can be classified into the following two categories. The first category is based on whether RXs are required to perform both Information Decoding (ID) and EH (co-located ID and EH)~\cite{MU-MISO-TBF-PS,MUTxMIMO,TX-PS,CL17_MISO_SWIPT,EBF2,MU_MISO_Access2,MU_MISO_Access,Close} or just act as ID or EH RXs (separated ID and EH)~\cite{MISO-Sep,EBF1,Rank1WCL,EH-BF}. The second category includes performance objectives like the minimization of TX power required for meeting Quality of Service (QoS) and EH constraints~\cite{MU-MISO-TBF-PS,MUTxMIMO,CL17_MISO_SWIPT,EBF2,MU_MISO_Access}, and throughput~\cite{TX-PS,MISO-Sep,MU_MISO_Access2,Close} or EH~\cite{Rank1WCL,EBF1,EH-BF} maximization for a given TX power budget and QoS constraints. {Recently in~\cite{K-Tier}, the impact of the density of small-cell base stations together with their transmit power and the time allocation factor between EH and information transfer was analytically investigated for $K$-tier heterogeneous cellular networks capable of SWIPT via TS. However, the jointly globally optimal TX precoding and RX PS operation for energy sustainable multiuser MISO SWIPT incorporating realistic nonlinear RF EH modeling is still unknown.} {Although the recent works~\cite{MU_MISO_Access,Close} considered  nonlinear RF EH modeling for studying multiantenna SWIPT systems, analytical investigations on the joint designs and their efficient algorithmic implementation were not provided.} {More specifically, \cite{MU_MISO_Access} considered that the harvested DC power is a known requirement at each EH-enabled RX, whereas~\cite{Close} focused on a point-to-point multiantenna scenario.}

\subsection{Paper Organization and Notations}
Section~\ref{sec:motiv} outlines the motivation and key contributions of this work. The considered system model description is presented in Section~\ref{sec:system_model}, while Section~\ref{sec:problem} details the proposed joint TX precoding and IoT PS optimization framework. Section~\ref{sec:optimal} discusses the optimal TX precoding for the considered energy sustainable IoT problem formulation, and the Global Optimization Algorithm (GOA) along with analytical bounds for the unique global optimum are presented in Section~\ref{sec:GOA}. Tight closed form approximations for the optimal TX Power Allocation (PA) and RX PS ratios are presented in Section~\ref{sec:approx}. A detailed numerical investigation of the proposed joint design along with the extensive performance comparisons against the relevant techniques is carried out in Section~\ref{sec:results}. The concluding remarks are mentioned in Section~\ref{sec:conclusion}.

Vectors and matrices are denoted by boldface lowercase and capital letters, respectively. The Hermitian transpose and trace of $\mathbf{A}$ are denoted by  $\mathbf{A}^{\rm H}$ and $\mathrm{tr}\left(\mathbf{A}\right)$, respectively, and $\mathbf{I}_{n}$ represents the $n \hspace{-0.25mm}\times \hspace{-0.25mm}n$ identity matrix ($n\geq2$). $\mathbf{A}^{-1}$ and $\mathbf{A}^{\frac{1}{2}}$ denote the inverse and square root, respectively, of a square matrix $\mathbf{A}$, whereas $\mathbf{A}\succeq0$ means that $\mathbf{A}$ is positive semi-definite. $\lVert\,\cdot\,\rVert$ and $\left|\,\cdot\,\right|$ are respectively used to represent the Euclidean norm of a complex vector and the absolute value of a complex scalar. $\mathbb{C}$ and $\mathbb{R}$ represent the complex and real number sets, respectively, and $\ceil{x}$ denotes the smallest integer larger than or equal to $x$.

\section{Motivation and Key Contributions}\label{sec:motiv} \color{black}
In this paper we are interested in the energy sustainability of IoT systems comprising of low power and data hungry network nodes capable of EH functionality. Since the lifetime of an EH IoT system~\cite{Lifetime} depends on the time elapsed until the first EH network node runs out of energy, maximizing the minimum (max-min) energy that can be harvested among the nodes is critical. Focusing on a MISO SWIPT multicasting IoT system where a multiantenna TX is responsible for simultaneously transferring information and power to low power and data hungry EH PS RXs, we study the EH fairness maximization problem. {Our proposed design aims at confronting the short wireless energy transfer range~\cite{IoT-Mag,ComMag} of the considered multicasting system by efficient utilization of the multiple TX antennas, thus, increasing the lifetime of RF EH IoT with individual QoS constraints.} In our optimization formulation, we consider a generic RF EH model for the IoT nodes that captures the nonlinear relationship between the harvested DC power and the received RF power for any practically available RF EH circuit~\cite{Powercast,RFEH2008,NonRFH,ICC17Wksp}. Moreover, we consider the general case of individual QoS requirements for the IoT nodes, which are represented by respective Signal-to-Interference-plus-Noise Ratio (SINR) constraints. Our goal is to jointly design TX precoding and RX PS in order to maximize the minimum harvested energy among RXs, while satisfying their individual QoS constraints. 

{The EH fairness problem has been recently investigated for secure MISO SWIPT systems~\cite{MaxMinSecure}. However, the existing TX precoding designs~\cite{MIMO_SWIPT,MU-MISO-TBF-PS,MUTxMIMO,EBF2,CL17_MISO_SWIPT,MU_MISO_Access2,TX-PS,MISO-Sep,EBF1,Rank1WCL,EH-BF} adopted an oversimplified linear RF EH model which has been lately shown~\cite{NonRFH,MU_MISO_Access,Close} to be incapable of capturing the operational characteristics of the available RF EH circuits.} In addition, the designs in~\cite{MIMO_SWIPT,MU-MISO-TBF-PS,MUTxMIMO,EBF2,CL17_MISO_SWIPT,MU_MISO_Access,MU_MISO_Access2,TX-PS,Close,MISO-Sep,EBF1,Rank1WCL,EH-BF} are based either on numerical solutions or iterative algorithms. Proofs of global optimality or analytical solutions shedding insights on the interplay between different system parameters are in general missing. Motivated by these observations, in this paper we present an efficient algorithm for obtaining the jointly globally optimal TX precoding and IoT PS design for the considered optimization objective, and provide explicit analytical insights on the presented design parameters. The key distinctions of this work compared with the state-of-the-art are: the design objective that incorporates practical energy fairness IoT demands, a novel solution methodology taking into account the nonlinearity of RF-to-DC rectification operation, and the nontrivial analytical insights on the joint solution that eventually result in efficient low complexity sub-optimal designs.
  
{Next we summarize the  novel contributions of this work.}
\begin{itemize}
	\item We first present our novel EH fairness maximization problem for energy sustainable MISO SWIPT multicasting IoT systems, while incorporating the practical nonlinear RF-to-DC rectification process. This nonconvex optimization problem is then transformed to an equivalent Semi-Definite Relaxation (SDR) formulation and we prove that it possesses a unique global optimum. 	
	\item We derive analytical tight upper and lower bounds for the global optimal value of the considered optimization problem. Capitalizing on these bounds, we then present an iterative GOA for the computation of the jointly globally optimal TX precoding and IoT PS ratios design. The fast convergence of the proposed algorithm to the global optimum of the targeted problem has been both analytically described and numerically validated.
	\item We present analytical insights for the optimal TX beamforming directions by investigating the interplay between the directions for either solely optimizing EH performance or the ID one. Tight analytical approximations for the optimal TX PA and uniform PS ratio at each IoT node for a given TX precoding design are also derived. {The latter insights and approximations have been used for designing two low complexity sub-optimal algorithms. These low complexity designs are suitable for low power IoT nodes and exhibit performance sufficiently close to the optimum one for certain cases of practical interest.}
	\item Our numerical results gain insights on the impact of key system parameters on the trade-off between optimized received RF power for EH at each node and their individual QoS requirements. We also carry out extensive comparative numerical investigations between our presented designs and relevant benchmark schemes which corroborate the utility of our optimization framework. 
\end{itemize}
The key challenges addressed in this paper for the considered MISO SWIPT multicasting IoT systems are: (a) incorporation of nonlinear RF-to-DC rectification operation in the proposed jointly globally optimal TX precoding and IoT PS ratio design; and (b) derivation of efficient (sufficiently close to the optimum for certain cases of practical interest) low complexity joint designs addressing the limited computational capability and energy constraints of low power IoT nodes.  
\color{black}


\section{System Description}\label{sec:system_model}
In this section we first present the considered MISO SWIPT multicasting IoT system together with the adopted channel model and underlying signal model. Then, we introduce the generic RF EH model under consideration that is capable of capturing the rectification operation of realistic RF EH circuits.

\subsection{System and Channel Models}
We consider a MISO SWIPT multicasting IoT system comprising of $K$ single-antenna IoT nodes and one sink node equipped with $N$ antennas that is responsible for simultaneously transferring information and power to the IoT nodes. Hereinafter, each $k$-th IoT node is denoted by RX$_k$ $\forall k\in\mathcal{K}\triangleq \{1,2,\ldots,K\}$ and the sink node is termed for simplicity TX. The multiantenna TX adopts Space Division Multiple Access (SDMA) with linear precoding according to which each RX$_k$ is assigned a dedicated precoding vector (or beam) for SWIPT. We denote by $s_k\in\mathbb{C}$ $\forall k\in\mathcal{K}$ the unit power data symbol at TX, which is chosen from a discrete modulation set and intended for RX$_k$. These $K$ data symbols are transmitted simultaneously through spatial separation with the aid of the $K$ linear precoding vectors $\mathbf{f}_1,\mathbf{f}_2,\ldots,\mathbf{f}_K\in\mathbb{C}^{N\times 1}$. As such, the complex baseband transmitted signal from the multiantenna TX is given by $\mathbf{x}\triangleq\textstyle\sum_{k=1}^{K}\mathbf{f}_ks_k$. For each precoding vector $\mathbf{f}_k$ associated with the data symbol $s_k$, we distinguish its following two components: \textit{i}) The phase part given by the normalized beamforming direction $\bar{\mathbf{f}}_k\triangleq\frac{\mathbf{f}_k}{\Vert\mathbf{f}_k\rVert}$; and \textit{ii}) The amplitude part representing the power $p_k$ allocated to $s_k$, i$.$e$.$, $p_k\triangleq\Vert\mathbf{f}_k\rVert^2$. Combining the latter two components of each $\mathbf{f}_k$ yields $\mathbf{f}_k=\sqrt{p_k}\bar{\mathbf{f}}_k$. For the transmitted signal $\mathbf{x}$, we assume that there exists a total power budget $P_T$, hence, it must hold $\textstyle\sum_{k=1}^{K}p_k\le P_T$.  

A frequency flat MISO fading channel is assumed for each of the $K$ wireless links that remains constant during one transmission time slot and changes independently from one slot to the next. We represent by $\mathbf{h}_k\in\mathbb{C}^{N\times 1}$ $\forall k\in\mathcal{K}$ the channel vector between the $N$-antenna TX and the single-antenna RX$_k$. The entries of each $\mathbf{h}_k$ are assumed to be independent Zero-Mean Circularly Symmetric Complex Gaussian (ZMCSCG) random variables with variance $\sigma_{h,k}^2$ that depends on the propagation losses of the TX to RX$_k$ transmission. The baseband received signal $y_k\in\mathbb{C}$ at RX$_k$ can be mathematically expressed as
\begin{equation}\label{Eq:System}
y_k \triangleq \mathbf{h}_{k}^{\rm H}\textstyle\sum_{j=1}^{K}\mathbf{f}_js_j + n_{a_k},
\end{equation}
where $n_{a_k}\in\mathbb{C}$ represents the zero-mean Additive White Gaussian Noise (AWGN) with variance $\sigma_{a_k}^2$. Assuming the availability of perfect Channel State Information (CSI) at both TX and RXs, we consider PS receptions~\cite{MIMO_SWIPT} according to which each RX$_k$ splits its received RF signal with the help of a power splitter. Particularly, an $\rho_k$ fraction of the received RF power at RX$_k$ is used for ID and the remaining $1-\rho_k$ fraction is dedicated for RF EH. Using this definition in \eqref{Eq:System}, the received signal available for ID at RX$_k$ is given by
\begin{equation}\label{eq:yi}
y_{k_i} \triangleq \sqrt{\rho_k}\left(\mathbf{h}_{k}^{\rm H}\textstyle\sum_{j=1}^{K}\mathbf{f}_js_j + n_{a_k}\right) + n_{d_k},
\end{equation}
where $n_{d_k}$ is a ZMCSCG distributed random variable with variance $\sigma_{d_k}^2$ representing the additional noise introduced during ID at RX$_k$. The resulting SINR for $s_k$ at each RX$_k$ can be derived as
\begin{eqnarray}\label{eq:SINR}
\text{SINR}_k \triangleq \frac{\rho_k \left|\mathbf{h}_{k}^{\rm H}\mathbf{f}_k\right|^2}{\rho_k\textstyle\sum_{j\in\mathcal{K}_k}\left|\mathbf{h}_k^{\rm H}\mathbf{f}_j\right|^2+\rho_k\sigma_{a_k}^2+\sigma_{d_k}^2},
\end{eqnarray}
where $\mathcal{K}_k\triangleq\mathcal{K}\setminus{k}.$ Similarly, the corresponding received signal available for RF EH at each RX$_k$ is given by 
\begin{equation}\label{eq:ye}
y_{k_e} \triangleq \sqrt{1-\rho_k}\left(\mathbf{h}_{k}^{\rm H}\textstyle\sum_{j=1}^{K}\mathbf{f}_js_j + n_{a_k}\right).
\end{equation}
Using the latter expression, the total received RF power at each RX$_k$ that is available for EH is defined as
\begin{equation}\label{eq:RP}
P_{R_k} \triangleq \left(1-\rho_k\right)\left(\textstyle\sum_{j=1}^{K}\left|\mathbf{h}_{k}^{\rm H}\mathbf{f}_j\right|^2 + \sigma_{a_k}^2\right).
\end{equation} 

\begin{figure}[!t]
	\centering 
	\subfigure[]{{\includegraphics[width=1.65in]{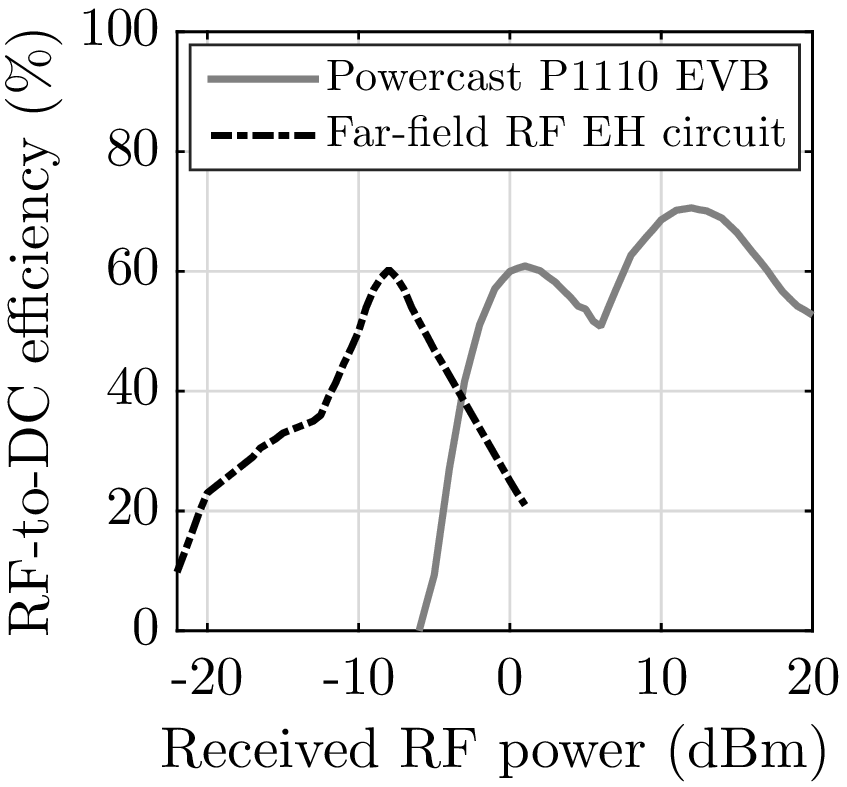} }} 
	\subfigure[]{{\includegraphics[width=1.65in]{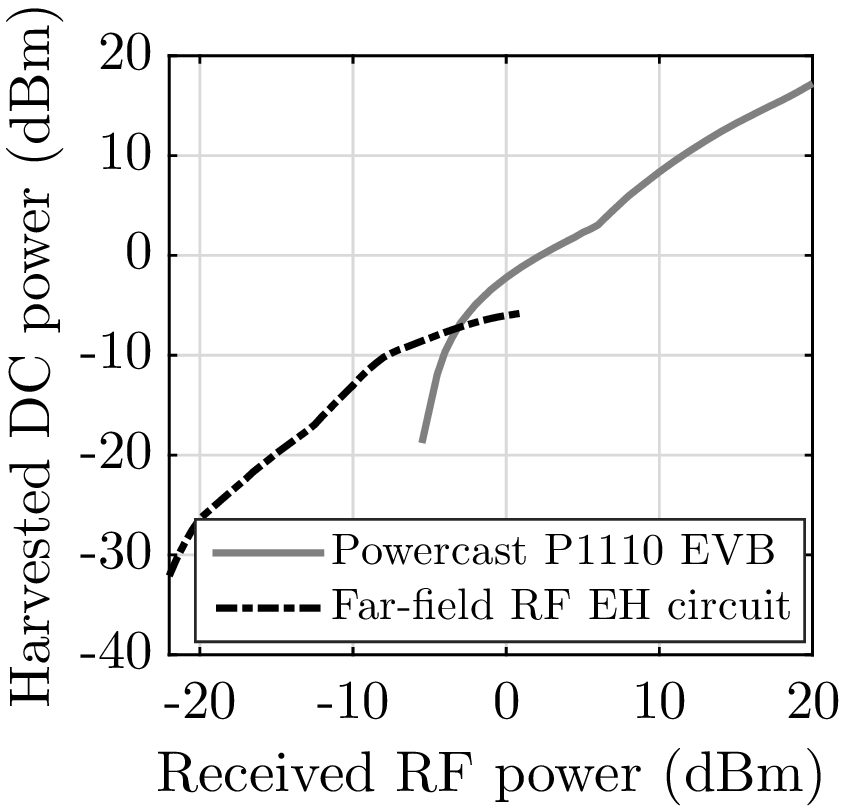} }}
	\caption{\small Variation of (a) RF-to-DC efficiency and (b) harvested DC power with the received RF power for practical RF EH circuit models.}
	\label{fig:RFEH} 
\end{figure} 

\subsection{RF Energy Harvesting Model}\label{sec:RFEH}
The harvested DC power at each RX$_k$ after RF-to-DC rectification of the received signal $y_{k_e}$ is given using \eqref{eq:RP} by 
\begin{equation}\label{eq:HP}
P_{H_k}=\eta\left(P_{R_k}\right)P_{R_k},
\end{equation}
where $P_{R_k}$ represents the received RF power at RX$_k$ and $\eta\left(\cdot\right)$ denotes the RF-to-DC rectification efficiency function of the RF EH circuitry used at each of the $K$ RXs. In general, $\eta\left(\cdot\right)$ is a positive \textit{nonlinear} function of the received RF power available for RF EH~\cite{Powercast,RFEH2008,NonRFH,ICC17Wksp}. This function is plotted in Fig$.$~\ref{fig:RFEH}(a) for two real-world RF EH circuits, namely, the commercially available Powercast P1110 Evaluation Board (EVB)~\cite{Powercast} and the circuit designed in~\cite{RFEH2008} for low power far field RF EH. It is obvious that the widely considered~\cite{MIMO_SWIPT,MU-MISO-TBF-PS,MUTxMIMO,EBF2,TX-PS,MISO-Sep,Rank1WCL,EBF1,EH-BF,MaxMinSecure} trivial linear RF EH model cannot efficiently describe practical rectification functionality, hence very recently, nonlinear models have been proposed~\cite{NonRFH,ICC17Wksp}. Despite the nonlinear relationship between the rectification efficiency and $P_{R_k}$, we note that due to the law of energy conservation holds that $P_{H_k}$ at each RX$_k$ is monotonically increasing with $P_{R_k}$, as shown in Fig$.$~\ref{fig:RFEH}(b). Hence, although the form of $\eta\left(\cdot\right)$ differs for different RF EH circuits, the non-decreasing nature of $P_{H_k}$ with $P_{R_k}$ is valid for all practical RF EH circuits~\cite{RFEH2008}. In other words, the relationship between the harvested DC power and received RF power can be defined as $P_{H_k}=\mathcal{F}\left(P_{R_k}\right)$, where $\mathcal{F}\left(\cdot\right)$ represents a \textit{nonlinear non-decreasing} function. We will exploit this feature in the following section including our proposed joint TX precoding and IoT PS optimization formulation.

\section{Optimization Problem Formulation}\label{sec:problem} 
We first present our energy sustainable IoT problem formulation and describe its key mathematical properties. Then, we present an equivalent SDR formulation for this problem and prove that it possesses a unique global optimum. We finally discuss the problem's feasibility conditions.

\subsection{Problem Definition}\label{sec:opt} 
We are interested in the joint design of TX precoding vectors $\{\mathbf{f}_k\}_{k=1}^K$ and RXs' PS ratios $\{\rho_k\}_{k=1}^K$ that maximizes the minimum of $\{P_{H_k}\}_{k=1}^K$ among the $K$ RF EH RXs, while satisfying all the underlying minimum SINR requirements $\bar{\gamma}_k$ $\forall k\in\mathcal{K}$ of all RXs. By using \eqref{eq:SINR}, \eqref{eq:HP}, and the total TX power $P_T$, the proposed optimization problem for the considered MISO SWIPT multicasting IoT system is formulated as 
\begin{eqnarray*}\label{eqOPT0}
	\begin{aligned} 
		\mathcal{OP}: &\underset{\{\mathbf{f}_k,\,\rho_k\}_{k=1}^K}{\text{max}}\,\textstyle\underset{k\in\mathcal{K}}{\text{min}}\,\; P_{H_k},\quad\text{s.t.:}\; ({\rm C1}): \text{SINR}_k\ge \bar{\gamma}_k,\,\forall k\in\mathcal{K},\\
		&({\rm C2}): \textstyle\sum_{k=1}^{K}\lVert\mathbf{f}_k\rVert^2\le P_T,\quad({\rm C3}): 0\le \rho_k\le 1,\forall k\in\mathcal{K},
	\end{aligned} 
\end{eqnarray*}
{where constraints $({\rm C1})$ and $({\rm C2})$ represent the minimum SINR requirements and maximum TX power budget, respectively. In addition, constraint $({\rm C3})$ includes the boundary conditions for $\rho_k$'s.} $\mathcal{OP}$ is a nonlinear nonconvex combinatorial optimization problem including the nonlinear function $\eta(\cdot)$ in the objective along with the coupled vectors $\{\mathbf{f}_k\}_{k=1}^K$ and ratios $\{\rho_k\}_{k=1}^K$ in both the objective and constraints. Specifically, quadratic terms of $\{\mathbf{f}_k\}_{k=1}^K$ appear in both the objective and constraints. To resolve these non tractable mathematical issues, we next present an equivalent SDR formulation for $\mathcal{OP}$ that can be solved optimally. Also, since RXs in the considered system are energy constrained, we assume that $\mathcal{OP}$ is solved at TX using the foreknown SINR demands $\{\bar{\gamma}_k\}_{k=1}^K$ along with the CSI knowledge of all involved links. After computing the optimal PS ratios, they are communicated to the corresponding RXs via appropriately designed control signals.  

\subsection{Semi-Definite Relaxation (SDR) Transformation}\label{subsec:gen_convx}
Using the definition $\mathbf{F}_k\triangleq\mathbf{f}_k\mathbf{f}_k^{\rm H}$ $\forall k\in\mathcal{K}$ in $\mathcal{OP}$ and ignoring the rank-$1$ constraint for each $\mathbf{F}_k$, an equivalent formulation $\mathcal{OP}1$ can be obtained after applying some algebraic rearrangements to the constraints and objective of $\mathcal{OP}$, as follows:
\begin{eqnarray*}\label{eqOPT1}
	\begin{aligned} 
		\mathcal{OP}1:&\underset{\mathcal{P},\{\mathbf{F}_k,\,\rho_k\}_{k=1}^K}{\text{max}}\quad\mathcal{P},\qquad\;\text{s.t.: }\;({\rm C3}),\\ 
		&({\rm C4}): \textstyle\sum_{j=1}^{K} \mathbf{h}_k^{\rm H}\mathbf{F}_j\mathbf{h}_k +\sigma_{a_k}^2\ge\frac{\mathcal{P}}{1-\rho_k},\forall k\in\mathcal{K},\\
		&({\rm C5}): \textstyle\frac{\mathbf{h}_k^{\rm H}\mathbf{F}_k\mathbf{h}_k}{\bar{\gamma}_k}-\!\textstyle\sum\limits_{j\in\mathcal{K}_k}\!\mathbf{h}_k^{\rm H}\mathbf{F}_j\mathbf{h}_k\ge\sigma_{a_k}^2+\frac{\sigma_{d_k}^2}{\rho_k},\,\forall k\in\mathcal{K},\\
		&({\rm C6}): \textstyle\sum_{k=1}^{K}\mathrm{tr}\left(\mathbf{F}_k\right)\le P_T, \quad\;({\rm C7}): \mathbf{F}_k\succeq0,\,\forall k\in\mathcal{K}.
	\end{aligned} 
\end{eqnarray*}
Constraints $({\rm C5})$ and $({\rm C6})$ represent the equivalent transformations for $({\rm C1})$ and $({\rm C2})$, respectively. {We have particularly replaced the TX precoding vectors $\{\mathbf{f}_k\}_{k=1}^K$  with their respective matrix definition $\{\mathbf{F}_k\}_{k=1}^K$. An additional variable $\mathcal{P}$ has been also included to reformulate the max-min $\mathcal{OP}$ problem to the simpler maximization problem $\mathcal{OP}1$ having $K$ additional constraints, as represented by the new constraint  $({\rm C4})$.} We have also replaced the harvested power maximization problem in $\mathcal{OP}$ with the corresponding received RF power for EH maximization in $\mathcal{OP}1$ by using the two key results as discussed next in Lemmas~\ref{lem:GCVX} and \ref{lem:equiv}. 
\begin{lemma}\label{lem:GCVX}
	\textit{The power $P_{R_k}=\left(1-\rho_k\right)\big(\textstyle\sum_{j=1}^{K} \mathbf{h}_k^{\rm H}\mathbf{F}_j\mathbf{h}_k+\sigma_{a_k}^2\big)$ at each RX$_k$ is jointly pseudoconcave in $\mathbf{F}_k$ and $\rho_k$.}
\end{lemma}
\begin{IEEEproof}
	The product of the two positive linear functions $\left(1-\rho_k\right)$ and $\left(\textstyle\sum_{j=1}^{K}\mathbf{h}_k^{\rm H}\mathbf{F}_j\mathbf{h}_k+\sigma_{a_k}^2\right)$ that defines the received RF power for EH at RX$_k$ is a pseudoconcave function~\cite[Tab$.$ 5.3]{avriel2010generalized}. This pseudoconcavity property holds jointly for $\mathbf{F}_k=\mathbf{f}_k\mathbf{f}_k^{\rm H}$ in the $P_{R_k}$ expression \eqref{eq:RP} and $\rho_k$.    
\end{IEEEproof}
\begin{lemma}\label{lem:equiv}
	\textit{The max-min problem of $\{P_{H_k}\}_{k=1}^{K}$ among the $K$ RXs is equivalent to the problem of maximizing the corresponding minimum received RF powers $\{P_{R_k}\}_{k=1}^{K}$.}
\end{lemma}
\begin{IEEEproof}
	From the discussion in Sec$.$~\ref{sec:RFEH} it follows that each harvested DC power $P_{H_k}$ is a non-decreasing function of the corresponding $P_{R_k}$. It also holds that the non-decreasing transformation of the pseudoconcave function $P_{R_k}$ is pseudoconcave~\cite{avriel2010generalized,boyd}. Using these properties together with Lemma~\ref{lem:GCVX}, we conclude that, since $P_{R_k}$ is jointly pseudoconcave in $\mathbf{F}_k$ and $\rho_k$, the same holds for $P_{H_k}$. In addition, it is known that a pseudoconcave function has a unique global maximum~\cite[Chap$.$ 3.5.9]{Baz}. Hence, maximizing the minimum among $\{P_{H_k}\}_{k=1}^{K}$ is equivalent to maximizing the minimum among $\{P_{R_k}\}_{k=1}^{K}$, and function $\eta\left(\cdot\right)$ defines the mathematical formula connecting their globally optimal solutions. 
\end{IEEEproof}

Using the latter two lemmas, we next prove the generalized convexity of $\mathcal{OP}1$ along with its equivalence to $\mathcal{OP}$.  
\begin{theorem}\label{th:GOP2}
	\textit{$\mathcal{OP}1$ having the unique globally optimal solution $\left(\mathcal{P}^*,\{\mathbf{F}_k^*,\,\rho_k^*\}_{k=1}^K\right)$ is an equivalent formulation for $\mathcal{OP}$.}
\end{theorem}
\begin{IEEEproof} 
	We first show that $\mathcal{OP}1$ belongs to the special class of generalized convex problems~\cite[Chapter 4.3]{Baz} that possess the unique global optimality property. Actually, $({\rm C3}),({\rm C6}),$ and $({\rm C7})$ in $\mathcal{OP}1$ are linear (i$.$e$.$, convex) constraints. Due to the linearity of the expression $\textstyle\frac{\mathbf{h}_k^{\rm H}\mathbf{F}_k\mathbf{h}_k}{\bar{\gamma}_k}-\textstyle\sum_{j\in\mathcal{K}_k} \mathbf{h}_k^{\rm H}\mathbf{F}_j\mathbf{h}_k$ and the convexity of $\rho_k^{-1}$ in $\mathbf{F}_k$ and $\rho_k$, $({\rm C5})$ is jointly quasiconvex. In addition, $({\rm C4})$ is jointly pseudoconcave from Lemma~\ref{lem:GCVX}. {Combining these properties of $\mathcal{OP}1$ constraints along with the linearity of $\mathcal{OP}1$ objective and result in~\cite[Theorem 4.3.8]{Baz}, yields that the Karush Kuhn Tucker (KKT) point of $\mathcal{OP}1$ is its globally optimal solution.}
	
It follows from Lemma~\ref{lem:equiv} that the harvested DC power max-min problem is equivalent to maximizing the minimum among the received RF powers. Using this result together with the epigraph transformation~\cite[Chap$.$ 4.2.4]{boyd} of $\mathcal{OP}$, we obtain $\mathcal{OP}1$ with an implicit rank-$1$ constraint to be satisfied by the globally optimal TX precoding matrix $\mathbf{F}_k^*$. As it will be proven in the following lemma, this condition is always implicitly met. Hence, $\mathcal{OP}$ and $\mathcal{OP}1$ are equivalent and the globally optimal solution $\{\mathbf{f}_k^*,\,\rho_k^*\}_{k=1}^K$ of $\mathcal{OP}$ can be obtained from the globally optimal solution $\left(\mathcal{P}^*,\{\mathbf{F}_k^*,\,\rho_k^*\}_{k=1}^K\right)$ of $\mathcal{OP}1$, where the optimal TX precoding vector $\mathbf{f}_k^*$ for each RX$_k$ is derived from the EigenValue Decomposition (EVD) of $\mathbf{F}_k^*$. 
\end{IEEEproof}

\begin{lemma}\label{lem:rank}
	\textit{The optimal solution $\left(\mathcal{P}^*,\{\mathbf{F}_k^*,\rho_k^*\}_{k=1}^K\right)$ of $\mathcal{OP}1$ implicitly satisfies the rank-$1$ condition for $\{\mathbf{F}_k^*\}_{k=1}^K$.}
\end{lemma}
\begin{IEEEproof}
	Keeping constraints $({\rm C3})$ and $({\rm C7})$ in $\mathcal{OP}1$ implicit and associating the Lagrange multipliers $\{\lambda_k\}_{k=1}^K$, $\{\mu_k\}_{k=1}^K$, and $\nu$, respectively, with the  constraints $({\rm C4})$, $({\rm C5})$, and $({\rm C6})$, the Lagrangian function of $\mathcal{OP}1$ is defined as
	\begin{align}\label{eq:lang1}
	\mathcal{L}&\big(\mathcal{P},\{\mathbf{F}_k,\rho_k,\lambda_k,\mu_k\}_{k=1}^K,\nu\big)\triangleq \textstyle\sum\limits_{k=1}^{K}\bigg(\mathrm{tr}\left(\mathbf{A}_k\mathbf{F}_k\right)-\mathrm{tr}\left(\mathbf{B}_k\mathbf{F}_k\right)\nonumber\\
	&\textstyle+\mathcal{P}+\nu P_T+\lambda_k\left(\sigma_{a_k}^2-\frac{\mathcal{P}}{1-\rho_k}\right)-\mu_k\left(\sigma_{a_k}^2+\frac{\sigma_{d_k}^2}{\rho_k}\right)\bigg)\hspace{-5mm}
	\end{align}
	with $\mathbf{A}_k\triangleq\left(\frac{\mu_k}{\bar{\gamma}_k}+\mu_k\right)\mathbf{h}_k \mathbf{h}_k^{\rm H}$ and $\mathbf{B}_k\triangleq\nu\,\mathbf{I}_{N}+\textstyle\sum\limits_{j=1}^{K}\left(\mu_k-\lambda_k\right)$ $\mathbf{h}_j \mathbf{h}_j^{\rm H}$. As $\mu_k\ge 0$ and $\bar{\gamma}_k>0$ $\forall k\in\mathcal{K}$, it holds $\mathbf{A}_k\succeq0$. Using this function, the dual function of $\mathcal{OP}1$ is given by
	\begin{eqnarray}\label{eq:dual}
	\textstyle g\left(\{\lambda_k,\mu_k\},\nu\right)\triangleq\!\!\underset{\substack{\mathcal{P},\{\mathbf{F}_k,\,\rho_k\}_{k=1}^K\\\{\mathbf{F}_k\succeq0,\,0<\rho_k<1\}}}{\text{max}}\!\!\mathcal{L}\left(\mathcal{P},\{\mathbf{F}_k,\rho_k,\lambda_k,\mu_k\},\nu\right)\!.\!\!\!\!
	\end{eqnarray}
	Since $\mathcal{OP}1$ has a unique globally optimal solution (cf$.$ Theorem~\ref{th:GOP2}), it holds from the strong duality principle~\cite[Section 6.2]{Baz} that its solution can be also obtained from the solution of the following dual problem:
	\begin{equation*}
	\mathcal{DP}1:\;\underset{\{\lambda_k\ge0,\mu_k\ge0\}_{k=1}^K,\,\nu\ge0}{\text{min}}\;g\left(\{\lambda_k,\mu_k\}_{k=1}^K,\nu\right).
	\end{equation*}
	Denoting the optimal solution of $\mathcal{DP}1$ as $\left(\{\lambda_k^*,\mu_k^*\}_{k=1}^K,\nu^*\right)$, the optimal power $\mathcal{P}^*$ and $\{\mathbf{F}_k^*,\rho_k^*\}_{k=1}^K$ that maximize the Lagrangian in \eqref{eq:lang1} is the optimal solution of $\mathcal{OP}1$. Since the variables $\{\mathbf{F}_k\}_{k=1}^K$ are decoupled from the remaining variables $\mathcal{P}$ and $\{\rho_k\}_{k=1}^K$ as shown in \eqref{eq:lang1}, we can compute $\{\mathbf{F}_k^*\}_{k=1}^K$ by solving the following equivalent problem (note that constant terms have been discarded in this equivalent formulation):
	\begin{eqnarray}\label{eq:prob2b}
	&\textstyle \underset{\{\widetilde{\mathbf{F}}_k\succeq0\}_{k=1}^K}{\text{max}}\mathrm{tr} \left((\widetilde{\mathbf{A}}_k^*)^{\rm H}\widetilde{\mathbf{F}}_k\widetilde{\mathbf{A}}_k^*\right)-\mathrm{tr}(\widetilde{\mathbf{F}}_k).
	\end{eqnarray}
	In \eqref{eq:prob2b}, $\widetilde{\mathbf{F}}_k\triangleq\left(\mathbf{B}_k^*\right)^{\frac{1}{2}}\mathbf{F}_k\left(\mathbf{B}_k^*\right)^{\frac{1}{2}}$ and $\widetilde{\mathbf{A}}_k^*\triangleq\left(\mathbf{A}_k^*\right)^{\frac{1}{2}}\left(\mathbf{B}_k^*\right)^{-\frac{1}{2}}$ are obtained by substituting the optimal solutions of $\mathcal{DP}1$ into $\mathbf{A}_k$ and $\mathbf{B}_k$, respectively. Here we have implicitly used $\mathbf{A}_k^*\succeq0$ and $\mathbf{B}_k^*\succeq0$, where the latter is imposed in order to have a bounded solution for $\mathcal{DP}1$. Using these properties along with the results proved in~\cite[Prop$.$ 1]{MUTxMIMO} (or~\cite[Th$.$ 1]{Rank1WCL}), the rank-$1$ property of the optimal solution $\widetilde{\mathbf{F}}_k^*$ of \eqref{eq:prob2b} can be shown by contradiction. Hence, each $\mathbf{F}_k^*=\left(\mathbf{B}_k^*\right)^{-\frac{1}{2}}\widetilde{\mathbf{F}}_k^*\left(\mathbf{B}_k^*\right)^{-\frac{1}{2}}$ has to be a rank-$1$ matrix like $\widetilde{\mathbf{F}}_k^*$ $\forall k\in\mathcal{K}$.
\end{IEEEproof}

\begin{remark}
{The outcomes of our energy sustainable IoT problem formulation that focuses on the practical QoS-aware harvested power fairness maximization are different from the objectives in existing multiuser SWIPT works~\cite{MU-MISO-TBF-PS,MUTxMIMO,EBF2,CL17_MISO_SWIPT,MU_MISO_Access,MU_MISO_Access2,TX-PS,Close,MISO-Sep,EBF1,Rank1WCL,EH-BF}.} This will be shown analytically in Section~\ref{sec:optimal} and through numerical validations in Section~\ref{sec:results}. The same holds for the joint TX precoding and IoT PS design of the proposed optimization problem that will be presented in the following sections.   
\end{remark}

\subsection{Feasibility Conditions}\label{sec:feas}
The feasibility of $\mathcal{OP}$ depends on the underlying SINR constraints $\{\bar{\gamma}_k\}_{k=1}^K$ of all $K$ RXs that need to be simultaneously met for a given total TX power budget $P_T$. To check whether $\{\bar{\gamma}_k\}_{k=1}^K$ can be satisfied, we solve the following problem:
\begin{eqnarray*}\label{eqOPT6}
	\begin{aligned} 
		\mathcal{OP}2:\;&\underset{\{\mathbf{f}_k\}_{k=1}^K}{\text{min}}\,\textstyle\sum_{k=1}^{K}\lVert\mathbf{f}_k\rVert^2,\qquad\text{s.t.: }\\
		&\textstyle({\rm C8}): \frac{\textstyle\left|\mathbf{h}_k^{\rm H}\mathbf{f}_k\right|^2}{\textstyle\sum_{j\in\mathcal{K}_k}\left|\mathbf{h}_k^{\rm H}\mathbf{f}_j\right|^2+\sigma_{a_k}^2 +\sigma_{d_k}^2}\ge \bar{\gamma}_k,\,\forall k\in\mathcal{K}.
	\end{aligned} 
\end{eqnarray*}
$\mathcal{OP}2$, which does not consider EH (i$.$e$.$, $\rho_k=1$ for each RX$_k$), has been widely studied and its globally optimal solution denoted by $\{\mathbf{f}_{k_I}\}_{k=1}^K$ is given by~\cite[eq$.$ (10)]{SPMag}. If $\textstyle\sum_{k=1}^{K}\lVert\mathbf{f}_{k_I}\rVert^2\le P_T$, then both $\mathcal{OP}$ and $\mathcal{OP}1$ are feasible, otherwise they are not. Also, to ensure $P_{H_k}^*>0$ $\forall k\in\mathcal{K}$ in $\mathcal{OP}$, $\mathcal{P}^*=\textstyle\underset{k\in\mathcal{K}}{\text{min}}\;\{P_{R_k}^*\}$ needs to satisfy $\mathcal{P}^*\ge \mathcal{S}_E$, where $\mathcal{S}_E$ is the receive energy sensitivity of the RF EH circuit~\cite{ComMag,RFEH2008}. 

\section{Optimal TX Precoding Design}\label{sec:optimal}  
Here we provide insights on the optimal TX precoding design for our energy sustainable IoT problem. These insights will be used later for implementing efficient algorithms for the jointly global optimal TX precoding and IoT PS design.  

\subsection{Optimal TX Precoding Structure}
The Lagrangian function $\mathcal{L}$ of $\mathcal{OP}1$ given by \eqref{eq:lang1} can be rewritten in terms of the precoding vectors $\{\mathbf{f}_k\}_{k=1}^K$ as
\begin{align}\label{eq:lang1b}
\mathcal{L}&\big(\mathcal{P},\{\mathbf{f}_k,\rho_k,\lambda_{k},\mu_{k}\},\nu\big)= \mathcal{P}+ \nu \left(P_T-\textstyle\sum_{k=1}^{K} \lVert\mathbf{f}_k\rVert^2\right)\nonumber\\
&+\textstyle\sum_{k=1}^{K}\lambda_{k}\Big(\textstyle\sum_{j=1}^{K}\left|\mathbf{h}_k^{\rm H}\mathbf{f}_j\right|^2 +\sigma_{a_k}^2-\frac{\mathcal{P}}{1-\rho_k}\Big)\nonumber\\
&+\textstyle\sum_{k=1}^{K}\mu_{k}\bigg(\textstyle\frac{\left|\mathbf{h}_k^{\rm H}\mathbf{f}_k\right|^2}{\bar{\gamma}_k}-\textstyle\sum\limits_{j\in\mathcal{K}_k}\!\!\left|\mathbf{h}_k^{\rm H}\mathbf{f}_j\right|^2-\sigma_{a_k}^2-\frac{\sigma_{d_k}^2}{\rho_k}\bigg).\hspace{-6mm}
\end{align} 
Then, from Theorem~\ref{th:GOP2}, each optimal $\mathbf{f}_k^*$ can be derived by solving $\frac{\partial \mathcal{L}}{\partial \mathbf{f}_k}=0$ (KKT condition for optimal TX precoding vector), which after few algebraic manipulations simplifies to 
\begin{eqnarray}\label{eq:fk2}
\textstyle\Big(\mathbf{I}_{N}+\sum\limits_{j=1}^K\left(\frac{\mu_{j}-\lambda_{j}}{\nu}\right) \mathbf{h}_j \mathbf{h}_j^{\rm H}\Big)\mathbf{f}_k= \left(\frac{1}{\bar{\gamma}_k}+1\right)\frac{\mu_{k}\mathbf{h}_k \mathbf{h}_k^{\rm H}\mathbf{f}_k}{\nu}.
\end{eqnarray}
Since $\left(\frac{1}{\bar{\gamma}_k}+1\right)\frac{\mu_{k}}{\nu} \mathbf{h}_k^{\rm H}\mathbf{f}_k$ in \eqref{eq:fk2} is a scalar, the optimal beamforming direction $\bar{\mathbf{f}}_k^*$ for each RX$_k$ can be obtained as
\begin{eqnarray}\label{eq:fk3}
&\bar{\mathbf{f}}_k^*=\frac{\left(\mathbf{I}_{N}+\sum\limits_{j=1}^K\left(\frac{\mu_{j}-\lambda_{j}}{\nu}\right) \mathbf{h}_j \mathbf{h}_j^{\rm H}\right)^{-1}\mathbf{h}_k}{\left\lVert\left(\mathbf{I}_{N}+\sum\limits_{j=1}^K\left(\frac{\mu_{j}-\lambda_{j}}{\nu}\right) \mathbf{h}_j \mathbf{h}_j^{\rm H}\right)^{-1}\mathbf{h}_k\right\rVert}.
\end{eqnarray}
The Lagrange multipliers $\lambda_{k}$ and $\mu_{k}$ in \eqref{eq:fk3} respectively correspond to the constraints $({\rm C4})$ and $({\rm C5})$, and are respectively related to the EH and SINR requirements for RX$_k$. When $\lambda_{k}=0$, $\bar{\mathbf{f}}_k^*$ coincides with the optimal TX precoding for ID (i$.$e$.$, no EH) as given by~\cite[eq$.$ (10)]{SPMag}. Whereas, $\lambda_{k}=\mu_{k}$ yields $\bar{\mathbf{f}}_k^*=\frac{\mathbf{h}_k}{\lVert \mathbf{h}_k\rVert}$, which refers to Maximal Ratio Transmission (MRT) for RX$_k$. Therefore, the structure of $\bar{\mathbf{f}}_k^*$ is a modified version of the regularized Zero Forcing (ZF) beamformer~\cite{EBF2} that balances the trade off between minimizing interference solely for efficient ID and maximizing the intended signal strength for efficient EH. 

\subsection{TX Precoding Design}\label{sec:weight}
As noted in the above discussion, the optimal TX beamforming direction $\{\bar{\mathbf{f}}_k^*\}_{k=1}^K$ needs to balance the trade-off between the beamforming directions intended for (i) maximizing the harvested energy fairness and (ii) the one targeting efficient information transfer by meeting the SINR demands with minimum required TX power budget. Capitalizing this insight we propose the following weighted TX beamforming direction: 
\begin{equation}\label{eq:wTXB}
\bar{\mathbf{f}}_{k_W} \triangleq \frac{w_k\bar{\mathbf{f}}_{k_I}+\left(1-w_k\right)\bar{\mathbf{f}}_{k_E}}{\left\lVert w_k\bar{\mathbf{f}}_{k_I}+\left(1-w_k\right)\bar{\mathbf{f}}_{k_E}\right\rVert},\,\forall\,k\in\mathcal{K},
\end{equation}
where $w_k\in\left(0,1\right)$ $\forall k\in\mathcal{K}$ represents the relative weight between the TX beamforming direction $\bar{\mathbf{f}}_{k_I}\triangleq\frac{\mathbf{f}_{k_I}}{\lVert \mathbf{f}_{k_I}\rVert}$ for efficient ID and the corresponding direction $\bar{\mathbf{f}}_{k_E}\triangleq\frac{\mathbf{f}_{k_E}}{\lVert \mathbf{f}_{k_E}\rVert}$ for efficient EH. We next derive the latter directions $k\in\mathcal{K}$  from their respective optimal TX precoding vectors $\{\mathbf{f}_{k_I}\}_{k=1}^K$ and $\{\mathbf{f}_{k_E}\}_{k=1}^K$, respectively.   
  
\subsubsection{Energy Fairness Maximization (EFM)}\label{sec:EHP}
By setting $\rho_k=0$ $\forall k\in\mathcal{K}$ (i$.$e$.$, no ID requirement at RXs) in $\mathcal{OP}1$, we focus solely on maximizing the EH fairness of the considered multicasting IoT system. For this setting, $\mathcal{OP}1$ reduces to the following EH fairness optimization problem:
\begin{eqnarray*}\label{eqOPT3}
\begin{aligned} 
\mathcal{OP}3:\;&\underset{\mathcal{P},\{\mathbf{F}_k\}_{k=1}^K}{\text{max}}\quad\mathcal{P},\qquad\;\text{s.t.: }\;({\rm C6}),\,({\rm C7}),\\ 
&({\rm C9}): \textstyle\sum_{j=1}^{K} \mathbf{h}_k^{\rm H}\mathbf{F}_j\mathbf{h}_k +\sigma_{a_k}^2\ge{\mathcal{P}},\forall k\in\mathcal{K}.
\end{aligned} 
\end{eqnarray*}
Since, $\mathcal{OP}3$ has a linear objective and constraints, it is convex. Let $\left(\mathcal{P}_E,\{\mathbf{F}_{k_E}\}_{k=1}^K\right)$ denote its jointly optimal solution. 
\begin{corollary}\label{cor:rankE}
	\textit{The optimal solution of $\mathcal{OP}3$ implicitly satisfies the rank-$1$ condition for $\{\mathbf{F}_{k_E}\}_{k=1}^K$. Hence, $\mathbf{f}_{k_E}$ for each RX$_k$ is derived from the EVD of $\mathbf{F}_{k_E}$.}
\end{corollary}
\begin{IEEEproof}
Keeping constraint $({\rm C7})$ in $\mathcal{OP}3$ implicit and associating the Lagrange multipliers $\{\lambda_{k_E}\}_{k=1}^K$ and $\nu_E$ with the constraints $({\rm C9})$ and $({\rm C6})$, respectively, the Lagrangian function of $\mathcal{OP}3$ is defined as
	\begin{align}\label{eq:lang3}
	\mathcal{L}_3\big(\mathcal{P},&\{\mathbf{F}_k,\lambda_{k_E}\}_{k=1}^K,\nu\big)\triangleq \mathcal{P}+\nu_E\left( P_T- \mathrm{tr}\left(\mathbf{F}_k\right)\right)\nonumber\\
	&+\textstyle\sum\limits_{k=1}^{K}\left(\mathrm{tr}\left(\mathbf{B}_{k_E}\mathbf{F}_k\right)+\lambda_{k_E}\left(\sigma_{a_k}^2-\mathcal{P}\right)\right),
	\end{align}
where $\mathbf{B}_{k_E}\triangleq\textstyle\lambda_{k_E}\sum_{j=1}^{K}\mathbf{h}_j \mathbf{h}_j^{\rm H}$. Following similar steps to the proof of Lemma~\ref{lem:rank}, the optimal precoding $\{\mathbf{F}_{k_E}\}_{k=1}^K$ can be obtained by solving the equivalent problem defined below:
	\begin{eqnarray}\label{eq:probEb}
	&\textstyle \underset{\{\widehat{\mathbf{F}}_k\succeq0\}_{k=1}^K}{\text{max}}\mathrm{tr} \left((\widehat{\mathbf{B}}_{k_E})^{\rm H}\,\widehat{\mathbf{F}}_k\,\widehat{\mathbf{B}}_{k_E}\right)-\mathrm{tr}(\widehat{\mathbf{F}}_k),
	\end{eqnarray}
where $\widehat{\mathbf{F}}_k\triangleq\left(\nu_E^*\right)^{\frac{1}{2}}\mathbf{F}_k\left(\nu_E^*\right)^{\frac{1}{2}}$ and $\widehat{\mathbf{B}}_{k_E}\triangleq\left(\mathbf{B}_k^*\right)^{\frac{1}{2}}\left(\nu_E^*\right)^{-\frac{1}{2}}$ are obtained by substituting the optimal solutions of the dual problem for $\mathcal{OP}3$ into $\mathbf{B}_k$ and $\nu_E$. Lastly, using $\widehat{\mathbf{B}}_{k_E}\succeq0$ and $\widehat{\mathbf{F}}_k^*\succeq0$ $\forall$ $k$ along with the results in Lemma~\ref{lem:rank}, the rank-$1$ properties of the optimal solution $\{\widehat{\mathbf{F}}_k^*\}_{k=1}^K$ of \eqref{eq:probEb}, and thus that of $\{\mathbf{F}_{k_E}\}_{k=1}^K$ in $\mathcal{OP}3$, can be shown by contradiction. 
\end{IEEEproof}
\begin{remark}
As it will be demonstrated in the numerical results of Section~\ref{sec:results} (cf$.$ Fig$.$~\ref{fig:energy}), the TX precoding design $\{\mathbf{f}_{k_E}\}_{k=1}^K$ obtained from the solution $\{\mathbf{F}_{k_E}\}_{k=1}^K$ of the EFM problem $\mathcal{OP}3$ outperforms the MRT design~\cite{EBF2} and TX energy beamforming design intended for maximizing the sum of harvested energies at all $K$ EH users~\cite{EBF1}.
\end{remark}

\subsubsection{Information Decoding (ID)}\label{sec:IDP}
When solely targeting TX precoding for enhancing the ID performance, we consider the case $\rho_k=1$ $\forall k\in\mathcal{K}$ (i$.$e$.$, no EH requirement at RXs). To derive  $\{\mathbf{f}_{k_I}\}_{k=1}^K$, we focus on solving $\mathcal{OP}2$ as defined in Section~\ref{sec:feas} that seeks for the precoding design minimizing the total TX power, while meeting the individual SINR requirements. Also, $\mathcal{OP}1$ is feasible only if $\textstyle\sum_{k=1}^{K}\lVert\mathbf{f}_{k_I}\rVert^2\le P_T$.

In the following section we present an iterative GOA for $\mathcal{OP}1$ that utilizes the optimal TX precoding vectors $\{\mathbf{f}_{k_I}\}_{k=1}^K$ and $\{\mathbf{f}_{k_E}\}_{k=1}^K$.

\section{Joint TX Precoding and RX Power Splitting}\label{sec:GOA}
Although $\mathcal{OP}1$ exhibits generalized convexity as shown in Section~\ref{subsec:gen_convx}, standard optimization tools (e$.$g$.$, the CVX Matlab package~\cite{cvx}) cannot be used due to the fact that constraint $({\rm C4})$ does not satisfy the Disciplined Convex Programming (DCP) rule set; this constraint includes the coupled term $\frac{\mathcal{P}}{1-\rho_k}$. To resolve this issue, we summarize in the sequel an iterative GOA for solving $\mathcal{OP}1$ that capitalizes on our derived tight upper and lower bounds for the optimal $\mathcal{P}^*$ of $\mathcal{OP}1$ and uses $\{\mathbf{f}_{k_E}\}_{k=1}^K$ and $\{\mathbf{f}_{k_I}\}_{k=1}^K$ of Section~\ref{sec:weight}.

\subsection{Tight Analytical Bounds for the Optimal $\mathcal{P}^*$ in $\mathcal{OP}1$} 
\subsubsection{Upper Bound $\mathcal{P}_{\rm ub}$ on $\mathcal{P}^*$}\label{sec:UB}
Clearly, the optimal solution $\mathcal{P}_E$ of $\mathcal{OP}3$ as defined in Section~\ref{sec:EHP}, provides an upper bound for $\mathcal{P}^*$ because there is no SINR constraint to be met. However, we next present a tighter upper bound that can be obtained from the solution of the following problem: 
\begin{eqnarray*}\label{eqOPT4}
	\begin{aligned} 
		\mathcal{OP}4:\;&\underset{\{\mathbf{F}_k,\,\rho_k\}_{k=1}^K}{\text{min}}\,\textstyle\sum_{k=1}^{K}\mathrm{tr}\left(\mathbf{F}_k\right),\qquad\text{s. t.: }\;({\rm C3}),\,({\rm C5}),\\ 
		&({\rm C7}),\,({\rm C10}): \textstyle\sum_{j=1}^{K} \mathbf{h}_k^{\rm H}\mathbf{F}_j\mathbf{h}_k +\sigma_{a_k}^2\ge\frac{\widehat{\mathcal{P}}}{1-\rho_k},\forall k\in\mathcal{K}.
	\end{aligned} 
\end{eqnarray*}
In $\mathcal{OP}4$, we seek for the minimum TX power required to meet $\widehat{\mathcal{P}}=\mathcal{P}_E$ together with the SINR demands $\{\bar{\gamma}_k\}_{k=1}^K$. The objective and constraints of this problem are jointly convex in $\{\mathbf{F}_k,\,\rho_k\}_{k=1}^K$ with $\{\mathbf{F}_k\}_{k=1}^K$ satisfying the rank-$1$ constraint. In addition, $\mathcal{OP}4$ satisfies the DCP rule set, hence, we can efficiently compute its jointly optimal solution $\left(\mathcal{P}_{4E},\{\mathbf{F}_{k_{4E}}\}_{k=1}^K\right)$ using~\cite{cvx}. The tight upper bound for $\mathcal{P}^*$ can thus be obtained as 
\begin{equation}\label{eq:Upper_Bound}
\mathcal{P}_{\rm ub}\triangleq\frac{\mathcal{P}_E P_T}{\sum_{k=1}^{K}\mathrm{tr}\left(\mathbf{F}_{k_{4E}}\right)}. 
\end{equation}
Note that, due to the presence of $\sigma_{a_k}^2,\sigma_{d_k}^2>0$ $\forall k\in\mathcal{K}$ in $(\rm C4)$ and $(\rm C5)$ along with the fact that $\mathcal{P}_E>\mathcal{P}^*$ and $\sum_{k=1}^{K}\mathrm{tr}\left(\mathbf{F}_{k_{4E}}\right)>P_T$, it holds that $\mathcal{P}_E>\mathcal{P}_{\rm ub}>\mathcal{P}^*$. 

\subsubsection{Lower Bound $\mathcal{P}_{\rm lb}$ on $\mathcal{P}^*$}\label{sec:LB}
With $\textstyle\sum_{k=1}^{K}\lVert\mathbf{f}_{k_I}\rVert^2\!\le\! P_T$, $\mathcal{OP}1$ is feasible and its solution $\mathcal{P}^*$ can be lower bounded as  
\begin{equation}\label{eq:P_I}
\mathcal{P}_I\triangleq\underset{k\in\mathcal{K}}{\text{min}}\,\left(1-\frac{1}{P_T}{\sum\limits_{j=1}^{K}\lVert\mathbf{f}_{j_I}\rVert^2}\right)\left(\sum\limits_{j=1}^{K}\left|\mathbf{h}_{k}^{\rm H}\mathbf{f}_{j_I}\right|^2 + \sigma_{a_k}^2\right). 
\end{equation}
To find a tighter lower bound, we then set $\widehat{\mathcal{P}}=\mathcal{P}_I$ in $\mathcal{OP}4$ and denote its jointly optimal solution by $\left(\mathcal{P}_{4I},\{\mathbf{F}_{k_{4I}}\}_{k=1}^K\right)$. The lower upper bound for $\mathcal{P}^*$ can be then derived using the solution of $\mathcal{OP}4$ as 
\begin{equation}\label{eq:Lower_Bound}
\mathcal{P}_{\rm lb}\triangleq\frac{\mathcal{P}_I P_T}{\sum_{k=1}^{K}\mathrm{tr}\left(\mathbf{F}_{k_{4I}}\right)}. 
\end{equation}
Lastly, since $\sigma_{a_k}^2,$ $\sigma_{d_k}^2>0$ and $\sum_{k=1}^{K}\mathrm{tr}\left(\mathbf{F}_{k_{4I}}\right)<P_T$, it yields $\mathcal{P}_I<\mathcal{P}_{\rm lb}<\mathcal{P}^*$.

Note that the tightness of the presented lower $\mathcal{P}_{\rm lb}$ and upper $\mathcal{P}_{\rm ub}$ bounds will be later numerically validated in Section~\ref{sec:results}.  

\subsection{Global Optimization Algorithm (GOA)}\label{sec:GoAc}
The proposed GOA for efficiently solving $\mathcal{OP}1$ is based on one-dimensional Golden Section Search (GSS) over the feasible range of $\mathcal{P}$ values, as given by the previously derived bounds $\mathcal{P}_{\rm lb}$ and $\mathcal{P}_{\rm ub}$. Its detailed algorithmic steps for the case where $\mathcal{OP}1$ is feasible are outlined in Algorithm~\ref{Algo:Opt}. Due to the generalized convexity of $\mathcal{OP}1$ and the tightness of $\mathcal{P}_{\rm lb}$ and $\mathcal{P}_{\rm ub}$, the proposed GOA converges fast to the optimal $\mathcal{P}^*$ satisfying the TX power budget expressed by constraint $(\rm C6)$ within an acceptable tolerance $\xi$. 
\begin{algorithm}[!t]
	{\small \caption{Global Optimization Algorithm (GOA) for $\mathcal{OP}1$}\label{Algo:Opt}
		\begin{algorithmic}[1]
			\Require Channel and system parameters $N,K,\{\mathbf{h}_k,\sigma^2_{a_k},\sigma^2_{d_k}\}_{k=1}^K,$ $\eta\left(\cdot\right),P_T,$ SINR demands $\{\bar{\gamma}_k\}_{k=1}^K,$ and tolerance $\xi$. 
			\Ensure Optimal TX precoding and PS ratios $\{\mathbf{f}_k^*,\rho_k^*\}_{k=1}^K$ for $\mathcal{P}^*$.
			\State Find $\mathcal{P}_{\rm ub}$ and $\mathcal{P}_{\rm lb}$ as in Sections$.$~\ref{sec:EHP} and~\ref{sec:IDP}.
			\State Set $\mathcal{P}_{\rm p}= \mathcal{P}_{\rm ub}-0.618\left(\mathcal{P}_{\rm ub}-\mathcal{P}_{\rm lb}\right)$.
			\State Set $\mathcal{P}_{\rm q}= \mathcal{P}_{\rm lb}+0.618\left(\mathcal{P}_{\rm ub}-\mathcal{P}_{\rm lb}\right)$.
			\State Solve $\mathcal{OP}4$ with $\widehat{\mathcal{P}}=\mathcal{P}_{\rm p}$ and store minimum TX power in $P_{T_{\rm p}}$\label{step:p}.
			\State Solve $\mathcal{OP}4$ with $\widehat{\mathcal{P}}=\mathcal{P}_{\rm q}$ and store minimum TX power in $P_{T_{\rm q}}$\label{step:q}.
			\State Set $\Delta=\min\{\left|P_T-P_{T_{\rm p}}\right|,\, \left|P_T-P_{T_{\rm q}}\right|\},$   and  $c=0$.
			\While{$\Delta>\xi$}
			\If{$\left|P_T-P_{T_{\rm p}}\right|\leq \left|P_T-P_{T_{\rm q}}\right|$}
			\State \!\!Set $\mathcal{P}_{\rm ub}=\mathcal{P}_{\rm q},\mathcal{P}_{\rm q}=\mathcal{P}_{\rm p},\mathcal{P}_{\rm p}= \mathcal{P}_{\rm ub}-0.618\left(\mathcal{P}_{\rm ub}-\mathcal{P}_{\rm lb}\right)$.
			\State \!\!Set $P_{T_{\rm q}}=P_{T_{\rm p}}$ and repeat step~\ref{step:p} to obtain $P_{T_{\rm p}}$.
			\Else
			\State \!\!Set $\mathcal{P}_{\rm lb}=\mathcal{P}_{\rm p},\mathcal{P}_{\rm p}=\mathcal{P}_{\rm q},\mathcal{P}_{\rm q}= \mathcal{P}_{\rm lb}+0.618\left(\mathcal{P}_{\rm ub}-\mathcal{P}_{\rm lb}\right)$.
			\State \!\!Set $P_{T_{\rm p}}=P_{T_{\rm q}}$ and repeat step~\ref{step:q} to obtain $P_{T_{\rm q}}$. 
			\EndIf
			\State Set $\Delta=\min\{\left|P_T-P_{T_{\rm p}}\right|,\, \left|P_T-P_{T_{\rm q}}\right|\}$ and $c=c+1$.
			\EndWhile
			\If{$\left|P_T-P_{T_{\rm p}}\right|\leq \left|P_T-P_{T_{\rm q}}\right|$}
			\State \!\!Set $\mathcal{P}^*=\mathcal{P}_{\rm p},$ repeat step~\ref{step:p} to obtain optimal $\{\mathbf{F}_k^*,\rho_k^*\}_{k=1}^K$.
			\Else
			\State \!\!Set $\mathcal{P}^*=\mathcal{P}_{\rm q},$ repeat step~\ref{step:q} to obtain optimal $\{\mathbf{F}_k^*,\rho_k^*\}_{k=1}^K$.
			\EndIf
			\State Obtain $\mathbf{f}_k^*$ using EVD of $\mathbf{F}_k^*$ $\forall k\in\mathcal{K}$.
		\end{algorithmic}
	} 
\end{algorithm} 

\subsubsection*{Complexity Analysis}
We now discuss the computational time required to obtain the joint TX precoding design and IoT PS ratios for $\mathcal{OP}1$ through GOA presented in Algorithm~\ref{Algo:Opt}. According to this algorithm, $\{\mathbf{f}_k^*,\rho_k^*\}_{k=1}^K$ are outputted when the resulting $\mathcal{P}^*$ is close up to the acceptable tolerance $\xi\ll1$ to $\mathcal{OP}1$'s globally optimal value. As seen from Algorithm~\ref{Algo:Opt}, the search space interval after each GSS iteration reduces by a factor of $0.618$~\cite[Chap$.$ 2.5]{GS}. This value combined with the quantity $\left(\mathcal{P}_{\rm ub}-\mathcal{P}_{\rm lb}\right)$ as the maximum search length for $\mathcal{P}^*$ gives the total number of iterations $c^* \triangleq \ceil[\Big]{\frac{\ln\left(\xi\right)-\ln\left(\mathcal{P}_{\rm ub}-\mathcal{P}_{\rm lb}\right)}{\ln\left(0.618\right)}}+1$ that are required for the termination of Algorithm~\ref{Algo:Opt}, while ensuring that the numerical error is less than $\xi$. Putting all together, we need to solve the problems $\mathcal{OP}2$ and $\mathcal{OP}3$ separately along with the $c^*$ runs for solving $\mathcal{OP}4$ to eventually obtain the jointly globally optimal solution of $\mathcal{OP}1$, and consequently $\mathcal{OP}$ due to equivalence. However, as will be numerically shown later on in Section~\ref{sec:results} (cf$.$ Fig$.$~\ref{fig:bounds}), since holds $\left(\mathcal{P}_{\rm ub}-\mathcal{P}_{\rm lb}\right)\ll1$, $c^*$ is generally very low in practice and corroborates the fast convergence of Algorithm~\ref{Algo:Opt}.
 
GOA provides an efficient way to obtain the joint TX precoding and IoT PS design for $\mathcal{OP}$, however, analytical insights on the jointly globally optimal parameters are difficult to be extracted. Recall that $\mathbf{f}_k=\sqrt{p_k}\bar{\mathbf{f}}_k$ $\forall$ $k\in\mathcal{K}$ and that analytical insights on each beamforming direction $\bar{\mathbf{f}}_k$ were presented in Section~\ref{sec:optimal}. We next present two sub-optimal designs that are based on the weighted TX beamforming directions given by \eqref{eq:wTXB} and exhibit low complexity computation of the weights $\{w_k\}_{k=1}^K$, the PA $\{p_k\}_{k=1}^K$, and IoT PS ratios $\{\rho_k\}_{k=1}^K$. It will be shown in the results later on that, for high values of the SINR demands, the sub-optimal algorithms perform sufficiently close to GOA, returning globally optimal solution.  

\section{Sub-optimal Precoding and Power Splitting}\label{sec:approx}
In this section we first present two jointly optimal PA and IoT PS schemes for given TX beamforming directions. The one assumes possibly different PS ratios among RXs and is termed as Dynamic Power Splitting (DPS), and the other considers Uniform Power Splitting (UPS). Capitalizing on these schemes, we then introduce two low complexity sub-optimal designs for the weights of the proposed weighted TX beamforming directions described in Section~\ref{sec:weight}.    

\subsection{Power Allocation (PA) and Dynamic Power Splitting (DPS)}\label{sec:DPS} 
Given the beamforming directions $\{\bar{\mathbf{f}}_k\}_{k=1}^K$ $\forall k\in\mathcal{K}$ and considering possibly different PS ratios among RXs, $\mathcal{OP}1$ reduces to the following joint TX PA and RX PS design problem: 
\begin{eqnarray*}\label{eqOPT5}
	\begin{aligned} 
		&\mathcal{OP}5:\;\underset{\mathcal{P},\{p_k,\,\rho_k\}_{k=1}^K}{\text{max}}\,\mathcal{P},\qquad\text{s.t.:}\quad({\rm C3}),\\
		&({\rm C11}): \textstyle\sum_{j=1}^{K} p_j\left|\mathbf{h}_k^{\rm H}\bar{\mathbf{f}}_j\right|^2+\sigma_{a_k}^2\ge \frac{\mathcal{P}}{1-\rho_k},\forall k\in\mathcal{K},\\
		&({\rm C12}): \frac{p_k\left|\mathbf{h}_k^{\rm H}\bar{\mathbf{f}}_k\right|^2}{\bar{\gamma}_k} -\textstyle\sum\limits_{j\in\mathcal{K}\setminus{k}}p_j\left|\mathbf{h}_k^{\rm H}\bar{\mathbf{f}}_j\right|^2-\sigma_{a_k}^2\ge\frac{\sigma_{d_k}^2}{\rho_k},\forall k\in\mathcal{K},\\
		&({\rm C13}): \textstyle\sum_{k=1}^{K}p_k\le P_T,\qquad({\rm C14}): p_k\ge 0,\forall k\in\mathcal{K}.
	\end{aligned} 
\end{eqnarray*}
The generalized convexity~\cite[Chapter 4.3]{Baz} of $\mathcal{OP}5$ can be proved in a similar fashion to $\mathcal{OP}1$. The objective of $\mathcal{OP}5$ is linear, constraints $({\rm C3})$, $({\rm C13})$, and $({\rm C14})$ are convex, and $({\rm C11})$ together with $({\rm C12})$ possess joint quasiconvexity in $\left(\mathcal{P},\{p_k,\,\rho_k\}\right)$. Based on this property, $\mathcal{OP}5$'s globally optimal solution can be obtained from the solution its KKT conditions. We thus associate the Lagrange multipliers $\{\lambda_{5_k}\}_{k=1}^K$, $\{\mu_{5_k}\}_{k=1}^K$, and $\nu_5$, respectively, with $({\rm C11})$, $({\rm C12})$, and $({\rm C13})$, while keeping $({\rm C3})$ and $({\rm C14})$ implicit. Hence, the Lagrangian $L_5$ of $\mathcal{OP}5$ is given by
\begin{align}
\hspace{-3mm}L_5&\big(\mathcal{P},\{p_k,\rho_k,\lambda_{5_k},\mu_{5_k}\},\nu_5\big)\triangleq \mathcal{P}+ \nu_5 \left[P_T-\textstyle\sum\limits_{k=1}^{K}p_k\right]\nonumber\\
&+\textstyle\sum\limits_{k=1}^{K}\lambda_{5_k}\Big[\textstyle\sum_{j=1}^{K} p_j\left|\mathbf{h}_k^{\rm H}\bar{\mathbf{f}}_j\right|^2+\sigma_{a_k}^2-\frac{\mathcal{P}}{1-\rho_k}\Big]\nonumber\\
&+\textstyle\sum\limits_{k=1}^{K}\mu_{5_k}\bigg[\textstyle\frac{p_k\left|\mathbf{h}_k^{\rm H}\bar{\mathbf{f}}_k\right|^2}{\bar{\gamma}_k}-\textstyle\sum\limits_{j\in\mathcal{K}\setminus{k}}\!\!p_j\left|\mathbf{h}_k^{\rm H}\bar{\mathbf{f}}_j\right|^2-\sigma_{a_k}^2-\frac{\sigma_{d_k}^2}{\rho_k}\bigg].\hspace{-3mm}
\end{align}
Together with constraints $({\rm C3}),({\rm C11})$--$({\rm C14})$ and the requirement for positive Lagrange multipliers, the KKT conditions for $\mathcal{OP}5$ are given by
\begin{subequations}
	\begin{align}\label{eq:KKT5a}
	\frac{\partial L_5}{\partial\mathcal{P}}=1-\textstyle\sum\limits_{k=1}^{K}\frac{\lambda_{5_k}}{1-\rho_k}=0,
	\end{align}
	\begin{align}\label{eq:KKT5b}
	\frac{\partial L_5}{\partial p_k}=&\;\textstyle\sum\limits_{j=1}^{K}\left|\mathbf{h}_j^{\rm H}\bar{\mathbf{f}}_k\right|^2\left(\lambda_{5_j}-\mu_{5_j}\right)+\mu_{5_k}\left|\mathbf{h}_k^{\rm H}\bar{\mathbf{f}}_k\right|^2\left(\frac{1}{\bar{\gamma}_k}+1\right)\nonumber\\
	&\;-\nu_5=0,\forall k\in\mathcal{K},
	\end{align}
	\begin{align}\label{eq:KKT5c}
	\frac{\partial L_5}{\partial\rho_k}=\frac{\mu_{5_k}\sigma_{d_k}^2}{\rho_k^2}-\frac{\lambda_{5_k}\mathcal{P}}{\left(1-\rho_k\right)^2}=0,\forall k\in\mathcal{K},
	\end{align}
	\begin{align}\label{eq:KKT5d}
	\nu_5 \left[P_T-\textstyle\sum_{k=1}^{K}p_k\right]=0,
	\end{align}
	\begin{align}\label{eq:KKT5e}
	\lambda_{5_k}\Big[\textstyle\sum\limits_{j=1}^{K} p_j\left|\mathbf{h}_k^{\rm H}\bar{\mathbf{f}}_j\right|^2+\sigma_{a_k}^2-\frac{\mathcal{P}}{1-\rho_k}\Big]=0,\,\forall k\in\mathcal{K},
	\end{align}
	\begin{align}\label{eq:KKT5f}
	\mu_{5_k}\bigg[\textstyle\frac{p_k\left|\mathbf{h}_k^{\rm H}\bar{\mathbf{f}}_k\right|^2}{\bar{\gamma}_k}-\textstyle\sum\limits_{j\in\mathcal{K}\setminus{k}}\!\!p_j\left|\mathbf{h}_k^{\rm H}\bar{\mathbf{f}}_j\right|^2-\sigma_{a_k}^2-\frac{\sigma_{d_k}^2}{\rho_k}\bigg]=0,\,\forall k\in\mathcal{K}.
	\end{align}
\end{subequations}

Since, power $P_{R_k}$ received at each RX$_k$ intended for EH is an increasing function of $P_T$, the TX power budget constraint $({\rm C13})$ is always satisfied at equality, thus causing $\nu_5>0$ due to complimentary slackness condition, as defined in~\eqref{eq:KKT5d}. We also observe from \eqref{eq:KKT5a}, \eqref{eq:KKT5b}, and \eqref{eq:KKT5c} that if $\nu_5>0$, then $\lambda_{5_k},\mu_{5_k}>0$ $\forall k\in\mathcal{K}$. Applying the latter result in \eqref{eq:KKT5f} yields
\begin{equation}\label{eq:KKT5f2}
\textstyle\frac{p_k\left|\mathbf{h}_k^{\rm H}\bar{\mathbf{f}}_k\right|^2}{\bar{\gamma}_k}-\textstyle\sum_{j\in\mathcal{K}\setminus{k}} p_j\left|\mathbf{h}_k^{\rm H}\bar{\mathbf{f}}_j\right|^2=\sigma_{a_k}^2+\frac{\sigma_{d_k}^2}{\rho_k},
\end{equation}
which can be rewritten in matrix form as follows
\begin{align}\label{eq:matr} 
\left[\begin{array}{c}p_1 \\p_2\\\vdots \\p_K\end{array}\right]=\mathbf{M}^{-1}\left[\begin{array}{c}\sigma_{a_1}^2+{\sigma_{d_1}^2}/{\rho_1} \\\sigma_{a_2}^2+{\sigma_{d_2}^2}/{\rho_2} \\\vdots\\ \sigma_{a_K}^2+{\sigma_{d_K}^2}/{\rho_K}\end{array}\right],
\end{align} 
where the elements of $\mathbf{M}\in\mathbb{R}_+^{K\times K}$ are defined as
\begin{equation}
[\mathbf{M}]_{ij}\triangleq\begin{cases}
\frac{1}{\bar{\gamma}_i}\left|\mathbf{h}_i^{\rm H}\bar{\mathbf{f}}_i\right|^2, & \text{$i=j$}\\
-\left|\mathbf{h}_i^{\rm H}\bar{\mathbf{f}}_j\right|^2, & \text{$i\neq j$}
\end{cases}. 
\end{equation}
By substituting \eqref{eq:matr} into \eqref{eq:KKT5e} and applying some mathematical simplifications, we obtain the following $K$ equations: 
\begin{equation}\label{eq:rhoP}
\textstyle\sum\limits_{j=1}^{K} \sum\limits_{i=1}^K[\mathbf{M}^{-1}]_{ji}\left(\sigma_{a_i}^2+\frac{\sigma_{d_i}^2}{\rho_i}\right)\left|\mathbf{h}_k^{\rm H}\bar{\mathbf{f}}_j\right|^2+\sigma_{a_k}^2=\frac{\mathcal{P}}{1-\rho_k},\forall k\in\mathcal{K}.
\end{equation}   
The optimal PS ratios and RF power $\mathcal{P}$ for EH, as respectively denoted by $\{\rho_k^*\}_{k=1}^K$ and $\mathcal{P}^*$, are finally obtained by solving a system of $K+1$ equations, particularly, the first $K$ equations of \eqref{eq:rhoP} together with the following equation:
\begin{equation}\label{eq:p-bud}
\textstyle\sum_{k=1}^K\sum_{j=1}^K [\mathbf{M}^{-1}]_{kj}\left(\sigma_{a_j}^2+\frac{\sigma_{d_j}^2}{\rho_j}\right)=P_T.
\end{equation}
The latter equation results from the substitution of \eqref{eq:matr} into $\sum_{k=1}^K p_k=P_T$. Since it holds $0 \le\rho_k^*\le1$ as well as $10^{-6}\le\mathcal{P}\le1$ (in W) due to wireless propagation characteristics and the low energy sensitivity of practical RF EH circuits (typically $\mathcal{S}_E\cong-23$dBm~\cite{RFEH2008}), the system of $K+1$ equations can be solved efficiently using commercial numerical solvers (like Matlab and Mathematica). This holds true due to the small search space the unknown parameters lie. Finally, the optimal PA $\{p_k^*\}_{k=1}^K$ is obtained by substituting $\{\rho_k^*\}_{k=1}^K$ into \eqref{eq:matr}.

\subsection{Power Allocation (PA) and Uniform Power Splitting (UPS)}\label{sec:UPS}
Given the beamforming directions $\{\bar{\mathbf{f}}_k\}_{k=1}^K$ $\forall k\in\mathcal{K}$ and considering UPS $\rho_k=\bar{\rho}$ $\forall k\in\mathcal{K}$ for all RXs, yields after substitution into \eqref{eq:p-bud}  
\begin{align}
\textstyle\sum_{k=1}^K\textstyle\sum_{j=1}^K[\mathbf{M}^{-1}]_{kj}\left(\sigma_{a_j}^2+\frac{\sigma_{d_j}^2}{\bar{\rho}}\right)=P_T.
\end{align}
The optimal UPS $\bar{\rho}^*$ is obtained from the latter equation as
\begin{align}\label{eq:UPSf}
\bar{\rho}^*=\frac{\textstyle\sum_{k=1}^K\textstyle\sum_{j=1}^K[\mathbf{M}^{-1}]_{kj}{\sigma_{d_j}^2}}{P_T-\textstyle\sum_{k=1}^K\textstyle\sum_{j=1}^K[\mathbf{M}^{-1}]_{kj}\sigma_{a_j}^2}.
\end{align}
Using this value in \eqref{eq:matr} and \eqref{eq:rhoP} the optimal PA $\{p_k^*\}_{k=1}^K$ and the optimal RF power $\mathcal{P}^*$ for EH are, respectively, given by
\begin{subequations}
	\begin{align}\label{eq:PA-U}
	p_k^*=\textstyle\sum_{j=1}^K[\mathbf{M}^{-1}]_{kj}\left(\sigma_{a_j}^2+\frac{\sigma_{d_j}^2}{\bar{\rho}^*}\right),\forall k\in\mathcal{K},
	\end{align}
	\begin{align}\label{eq:PS-U}
	\mathcal{P}^*= \left(1-\bar{\rho}^*\right)\left(p_1^*\left|\mathbf{h}_1^{\rm H}\bar{\mathbf{f}}_1\right|^2\left(\frac{1}{\bar{\gamma}_1}+1\right)-\frac{\sigma_{d_1}^2}{\bar{\rho}^*}\right).
	\end{align}
\end{subequations}  
Obviously, for this case of given TX beamforming directions and UPS, the jointly optimal PA and UPS design is obtained in closed form as defined in \eqref{eq:PA-U} and \eqref{eq:UPSf} with corresponding optimal RF power for EH as given by \eqref{eq:PS-U}.   

\subsection{Low Complexity Sub-optimal Designs}\label{sec:iteration}
We next present two iterative schemes for computing the weights $\{w_k\}_{k=1}^K$ of the weighted TX beamforming directions given by \eqref{eq:wTXB}, which together with the previous joint PA and IoT PS schemes comprise our two proposed low complexity sub-optimal designs for $\mathcal{OP}$. Their low complexity comes from the fact that, for given TX beamforming directions, the jointly optimal PA and IoT UPS is obtained in closed form as shown in Section~\ref{sec:UPS} (i$.$e$.$, using \eqref{eq:UPSf}, \eqref{eq:PA-U} and \eqref{eq:PS-U}) and in an efficient way as presented in Section~\ref{sec:DPS} (i$.$e$.$, by solving \eqref{eq:rhoP} and \eqref{eq:p-bud}) for the DPS case.  

\subsubsection{Uniform Weight Allocation (UWA)}\label{sec:UWs}
In the UWA scheme it is considered that $w_k=\bar{w}\in(0,1)$ $\forall$ $k\in\mathcal{K}$. Without loss of generality, we assume that the common $\bar{w}$ varies in $x$ discrete steps ranging from $0$ to $1$, resulting in the weight allocation $\left\lbrace0,\frac{1}{x-1},\frac{2}{x-1},\ldots,\frac{x-2}{x-1},1\right\rbrace$. To compute $\bar{w}^*$ yielding the maximum $\mathcal{P}$, one needs to evaluate $\mathcal{P}$ for all assumed $x$ allocations and then select the best among them.

\subsubsection{Distinct Weight Allocation (DWA)}\label{sec:DWs}
For this scheme we consider that each weight $w_k$ $\forall$ $k\in\mathcal{K}$ varies in $x$ discrete steps. Instead of performing $K$ dimensional traverses over the possible weight allocations that imposes increased complexity, we first sort the values $\{\lVert \mathbf{h}_k\rVert\}_{k=1}^K$ for all $K$ RXs. Then, we proceed by optimizing the weight for the RX having the lowest channel gain (i$.$e$.$, RX$_i$ for which $\hat{i}=\argmin_k\,\lVert \mathbf{h}_k\rVert$), while setting unit weights for all other RXs (i$.$e$.$, $w_k=1$ $\forall$ $k\neq \hat{i}$, which means that for these RXs ID is solely chosen). The optimization continues by selecting the weight that results in the highest $\mathcal{P}$ among the $x$ possible weight allocations for the current RX. At most $x K$ discrete weight allocations need to be checked till obtaining $\{w_k^*\}_{k=1}^K$ yielding the maximum $\mathcal{P}$. 

\section{Numerical Results and Discussion}\label{sec:results}
In this section we evaluate the presented joint TX precoding and IoT PS designs for the considered MISO SWIPT multicasting IoT system. In figures that follow we have set $P_T=10$W, $K=4$, $\sigma_{a_k}^2\!=\!-70$dBm, $\sigma_{d_k}^2\!=\!-50$dBm, $\mathcal{S}_E=-30$dBm, $\xi=10^{-4}$, $x=20$, and in certain cases $\bar{\gamma}_k=\bar{\gamma}$ $\forall k\in\{1,2,3,4\}$. In addition, $\sigma_{h,k}^2=\theta d_k^{-\alpha}$ with $\theta=0.1$ being the average channel attenuation at unit reference distance, $d_k$ is TX to RX$_k$ distance, and $\alpha=2.5$ is the path loss exponent. The $K$ RXs have been placed uniformly over a square field with length $L=\{5,6\}$m and the TX was placed at its center. For the average performance results included in the figures we have used $10^3$ independent channel realizations. 

\subsection{Energy Harvesting vs SINR Tradeoff}\label{sec:valid}
\begin{figure}[!t]
	\centering 
	{{\includegraphics[width=3.1in]{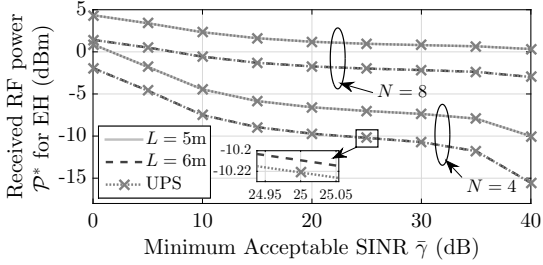} }}
	\vspace{-2mm}\caption{\small Received RF power $\mathcal{P}^*$ for EH in dBm as a function of the SINR $\bar{\gamma}$ in dB for $K=4$ and different values of $L$ and $N$.}
	\label{fig:tradeoff} 
\end{figure}

We first plot in Fig$.$~\ref{fig:tradeoff} the average optimal received RF power $\mathcal{P}^*$ for EH via GOA as a function of $\bar{\gamma}$ values in dB for different combinations of $L$ and $N$. This plot is also known as EH power versus SINR tradeoff. As shown, lower $L$ (i$.$e$.$, lesser propagation loss) and higher $N$ (i$.$e$.$, larger beamforming gain) values improve this tradeoff. It is also observed that as $\bar{\gamma}$ increases from $0$dB to $40$dB, there is a lower decrease of about $4$dBm in $\mathcal{P}^*$ for $N=8$ as compared to the decrease of $12$dBm for $N=4$. Recall that $K=4$ RXs have been considered. This corroborates the utility of having more TX antennas for improved EH power versus SINR tradeoff. In addition, for the case of field size $L=6$m, $\mathcal{P}^*$ is about $5$dBm lower than that for $L=5$m. 
Within this figure, we also sketch the obtained tradeoff for the sub-optimal design using UPS (i$.$e$.$, $\rho_k=\bar{\rho}$ $\forall k\in\{1,2,3,4\}$). This design performs very close to the sub-optimal DPS one that optimizes the individual PSs exhibiting lower complexity. Recall that with the UPS-based design the jointly optimal PA and UPS are obtained in closed form, and the TX beamforming weights are computed via a simple one-dimensional search.

\begin{figure}[!t]
	\centering 
	{{\includegraphics[width=3.1in]{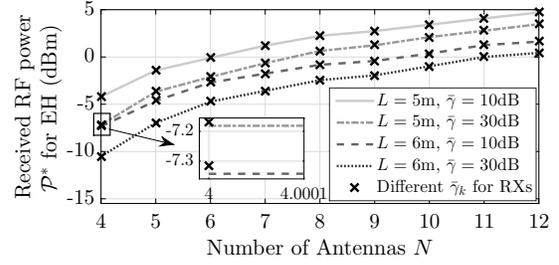} }}
	\vspace{-2mm}\caption{\small Received RF power $\mathcal{P}^*$ for EH in dBm as a function of the number of TX antennas $N$ for $K=4$ and different $L$ and $\bar{\gamma}$ $\left(\text{or }\{\bar{\gamma}_k\}_{k=1}^4\right)$ values.}
	\label{fig:antennaG}  
\end{figure}
The role of the number of TX antennas $N$ in $\mathcal{P}^*$ performance using GOA is depicted in Fig.~\ref{fig:antennaG} for different combinations of $L$ and $\bar{\gamma}$ (or $\{\bar{\gamma}_k\}_{k=1}^4$) values. Increasing $N$ from $4$ to $12$ improves $\mathcal{P}^*$ at each of the $K=4$ RXs by about $10$dBm. As expected due to the low energy transfer efficiency of SWIPT systems, the lower field size $L=5$m yields larger $\mathcal{P}^*$ at $\bar{\gamma}=30$dB as compared to that of $L=6$m at $\bar{\gamma}=10$dB. {For the case of unequal SINR demands at the $K=4$ RXs, we have used the values $\bar{\gamma}_1=8$, $\bar{\gamma}_2=9$, $\bar{\gamma}_3=11$, and $\bar{\gamma}_4=12$, with mean among them being the common SINR value $\bar{\gamma}=10$ or $\bar{\gamma}=10$dB. Likewise, for the common SINR being mean value $\bar{\gamma}=1000$ (or $30$dB), we have set $\bar{\gamma}_1=500$, $\bar{\gamma}_2=750$, $\bar{\gamma}_3=1250$, and $\bar{\gamma}_4=1500$.} Although a similar trend happens in both distinct SINR scenarios, it is noted that they both result in an average increase of about $0.32$\% in the average received RF power $\mathcal{P}^*$ for EH as compared to the scenario having the same SINR demands for all four RXs. 
\begin{figure}[!t]
	\centering 
	{{\includegraphics[width=3.1in]{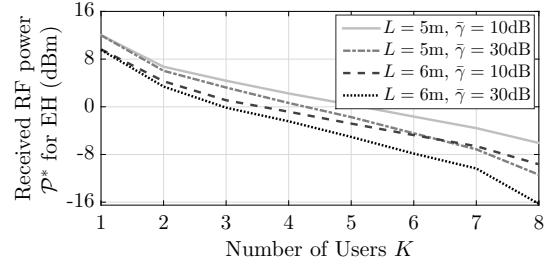} }} 
	\vspace{-2mm}\caption{\small Received RF power $\mathcal{P}^*$ for EH in dBm as a function of the RXs' number $K$ for $N=8$ and different combinations of $L$ and $\bar{\gamma}$.}
	\label{fig:users}
\end{figure} 
In Fig.~\ref{fig:users} we investigate the effect of IoT density for the parameter setting of Fig.~\ref{fig:antennaG} expect for assuming $N=8$ and varying the number of RXs $K$. It can be observed that $\mathcal{P}^*$ degrades significantly as the TX load to transfer energy to more RXs increases.   

\begin{figure}[!t]
	\centering\color{black}
	\subfigure[Rectification Efficiency.]{{\includegraphics[width=1.6in]{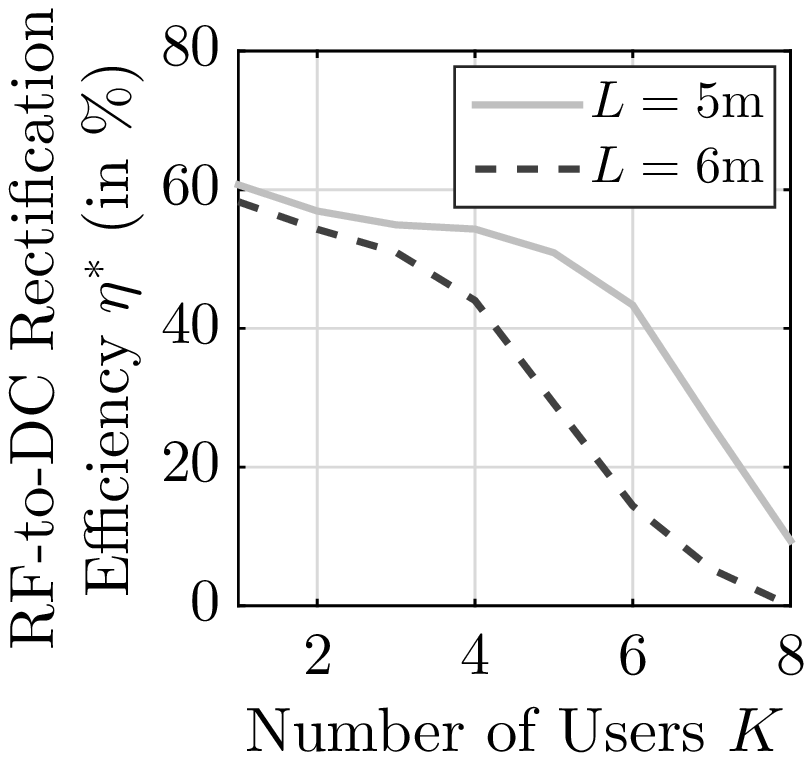} }}\;
	\subfigure[Harvested DC Power.]
	{{\includegraphics[width=1.6in]{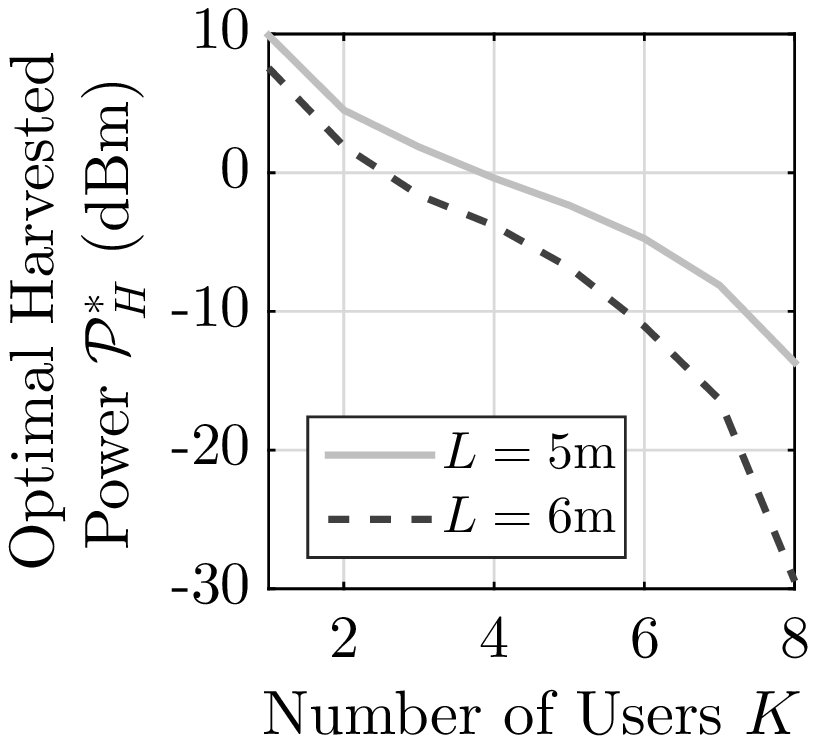} }} 
	\vspace{-2mm}\caption{\small Variation of the rectification efficiency $\eta^*$ and harvested power $\mathcal{P}_H^*$ for the EH circuit~\cite{Powercast} with varying $K$ for $N=8$, $\bar{\gamma}=10$dB, and $L=\{5,6\}$m. Underlying $\mathcal{P}^*$ variation is illustrated in Fig.~\ref{fig:users}.} 
	\label{fig:Eta} \color{black}
\end{figure}
{The impact of the nonlinear rectification efficiency $\eta$ on the optimized harvested DC power $\mathcal{P}_H^*\triangleq\eta^*\,\mathcal{P}^*$ with varying number of RXs is showcased in Fig.~\ref{fig:Eta} for the case where the RF EH unit of each RX is the Powercast P1110 EVB~\cite{Powercast}. The results for $\eta^*$ and $\mathcal{P}_H^*$, as respectively plotted in Figs.~\ref{fig:Eta}(a) and~\ref{fig:Eta}(b), are obtained using the relationship between $\mathcal{P}_H^*$ and $\mathcal{P}^*$ for the considered board, which has been analytically characterized by~\cite[eq. (6)]{ICC17Wksp}. Unlike the variation of $\eta^*$ with $K$, $\mathcal{P}_H^*$ follows a monotonically decreasing trend with increasing $K$, a trend that is actually very similar to the one followed by $\mathcal{P}^*$ in Fig.~\ref{fig:users}. This corroborates the discussion with respect to the claims made in Lemma~2 and the RF EH characteristics as plotted in Fig.~\ref{fig:RFEH}.}

\begin{figure}[!t]
	\centering 
	{{\includegraphics[width=3.1in]{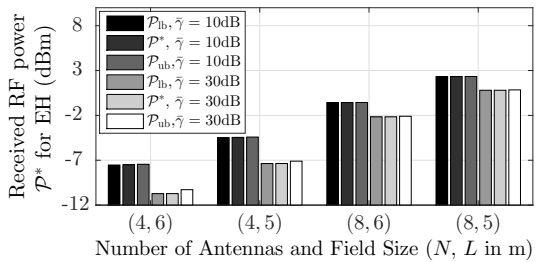} }} 
	\vspace{-2mm}\caption{\small Tightness of the lower $\mathcal{P}_{\rm lb}$ and upper $\mathcal{P}_{\rm ub}$ bounds on $\mathcal{P}^*$ for $K=4$ and different combinations of $\bar{\gamma}, L,$ and $N$ values.}
	\label{fig:bounds} 
\end{figure}
To corroborate the fast convergence of the proposed GOA in Algorithm~\ref{Algo:Opt}, we illustrate in Fig.~\ref{fig:bounds} the difference between our derived lower $\mathcal{P}_{\rm lb}$ and upper $\mathcal{P}_{\rm ub}$ bounds along with the optimal $\mathcal{P}^*$ for $K=4$ RXs and different values of $\bar{\gamma}, L,$ and $N$. It can be shown that the search space for $\mathcal{P}^*$ is very small (i$.$e$.$, $\mathcal{P}_{\rm ub}-\mathcal{P}_{\rm lb}\ll 1$). Particularly, the average difference between $\mathcal{P}_{\rm ub}$ and $\mathcal{P}_{\rm lb}$ is less than $0.004$mW (or $<-24$dBm) for $\bar{\gamma}=10$dB and less than $0.01$mW (or $<-20$dBm) for $\bar{\gamma}=30$dB. This fact validates our claims for the quality of our presented bounds for $\mathcal{P}^*$ and the fast convergence of GOA to the jointly globally optimal TX precoding and IoT PS design. 

\subsection{Optimal TX Beamforming Direction, PA, and RX PS Ratios}
We now focus on our two presented low complexity sub-optimal schemes in Section~\ref{sec:iteration} and investigate the derived designs for the TX beamforming directions and PA (combinedly forming the TX precoding design) as well as the RX PS ratios under different system parameter settings. 
\begin{figure}[!t]
	\centering 
	{{\includegraphics[width=3.1in]{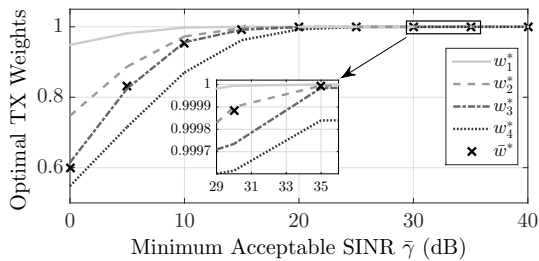} }}
	\vspace{-2mm}\caption{\small Optimal weights $\{w_1^*,w_2^*,w_3^*,w_4^*\}$ for the proposed weighted TX beamforming directions as a function of the SINR $\bar{\gamma}$ in dB for $N=K=4$ and $L=5$m.}
	\label{fig:precoding} 
\end{figure}
In Fig.~\ref{fig:precoding} we first plot the optimal weights $\{w_1^*,w_2^*,w_3^*,w_4^*\}$ versus $\bar{\gamma}$ in dB that are assigned to the weighted TX beamforming directions given by \eqref{eq:wTXB} using the proposed iterative DWA scheme. For this figure we have considered $N=K=4$ and $L=5$m. As shown, each weight increases with increasing SINR demand. This implies that the relative importance of TX precoding for efficient ID alone (as represented by $w_k^*\approx1$ $\forall$~$k\in\{1,2,3,4\}$) gets significantly higher than the precoding designed for maximizing the EH performance (as represented by $w_k^*\approx0$ $\forall$~$k\in\{1,2,3,4\}$). In addition, it can be seen that the weight values in this figure are mainly approaching their largest values (i$.$e$.$, greater than $0.5$ even for $\bar{\gamma}=0$dB). This shows that the designed TX beamforming directions approaching the optimal ones from GOA are closer to the ID-based TX precoding (i$.$e$.$, $\{\mathbf{f}_{k_I}\}_{k=1}^4$). 
\begin{figure}[!t]
	\centering 
	{{\includegraphics[width=3.1in]{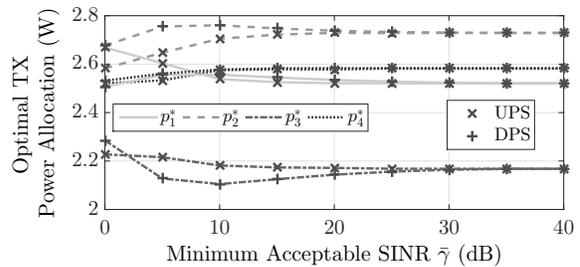} }} 
	\vspace{-2mm}\caption{\small Optimal TX PA $\{p_1^*,p_2^*,p_3^*,p_4^*\}$ with UPS and DPS as a function of the SINR $\bar{\gamma}$ in dB for $N=K=4$, $L=5$m, and the TX beamforming directions $\{\bar{\mathbf{f}}_{k_W}^*\}_{k=1}^4$.}
	\label{fig:Power}   
\end{figure}
We now use the parameter setting of Fig.~\ref{fig:precoding} and the derived TX beamforming directions to plot in Fig.~\ref{fig:Power} the variation of the optimal TX PA $\{p_1^*,p_2^*,p_3^*,p_4^*\}$ for both the DPS and UPS schemes. As shown for high SINR demands (i$.$e$.$, for $\bar{\gamma}\geq20$dB), $p_1^*,p_2^*,p_3^*,$ and $p_4^*$ for UPS and DPS closely match among each other. This trend again corroborates the fact that the adoption of the UPS scheme is a good approximation for MISO SWIPT multicasting IoT systems with high QoS constraints. It is also evident that for high SINR values the optimal PA becomes independent of the $\bar{\gamma}$ variations.
\begin{figure}[!t]
	\centering\vspace{-2mm}
	{{\includegraphics[width=3.1in]{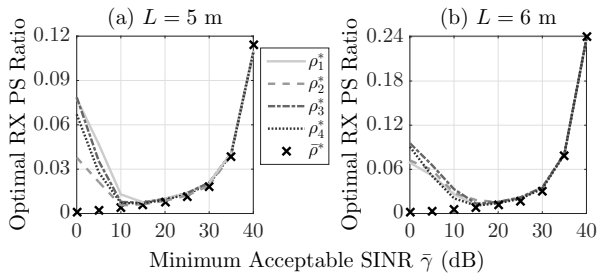} }} 
	\vspace{-2mm}\caption{\small Optimal DPS $\{\rho_1^*,\rho_2^*,\rho_3^*,\rho_4^*\}$ and UPS $\bar{\rho}^*$ ratios as a function of SINR $\bar{\gamma}$ in dB for $N=K=4$ under: (a) $L=5$m and (b) $L=6$m.}
	\label{fig:PS-ratio1}
\end{figure} 

The optimal PS ratios using both DPS and UPS schemes is illustrated in Fig.~\ref{fig:PS-ratio1} as a function of the SINR $\bar{\gamma}$ in dB for $N=K=4$ as well as $L=5$m and $6$m. For $\bar{\gamma}\ge20$dB the optimal PS ratios $\rho_1^*,\rho_2^*,\rho_3^*,$ and $\rho_4^*$ with DPS increase with increasing $\bar{\gamma}$. Interestingly, all ratios become nearly equal for $\bar{\gamma}\ge20$dB and match very closely with the optimal UPS ratio $\bar{\rho}^*$. This again showcases that the UPS-based scheme provides a very good approximation for the DPS one, especially for high QoS constraints. However, at the low SINR regime, the optimal DPS-based PS ratios follow a different trend from the UPS one. This has been also noticed in Figs.~\ref{fig:precoding} and~\ref{fig:Power} where power allocations and TX precoding for these two schemes were designed as different. 

\subsection{Comparisons with Relevant Designs}\label{sec:comp}
The proposed joint TX precoding and IoT PS design will be compared next with benchmark designs available in the relevant literature~\cite{MU-MISO-TBF-PS,EBF2,EBF1}. {As shown in the previous figures, our joint design based on the UPS scheme exhibits low complexity computation of the involved parameters  and performs sufficiently close to our optimal joint design obtained from GOA. This low computational overhead is achieved using the closed form expressions for PA and UPS, along with a simpler one-dimensional search for obtaining the TX beamforming weights.} We will thus consider this scheme in the performance comparisons that follow incorporating either the UWA or the DWA technique for the TX beamforming weight computation. We term these two versions of our joint design as Proposed-UWA-UPS and Proposed-DWA-UPS, respectively. For the benchmark designs we use the terminology SINR-UPS for the design in \cite{MU-MISO-TBF-PS}, as well as MRT-ZF-UWA-UPS and MRT-ZF-DWA-UPS for those in \cite{EBF2}.
\begin{figure}[!t]
	\centering 
	{{\includegraphics[width=3.1in]{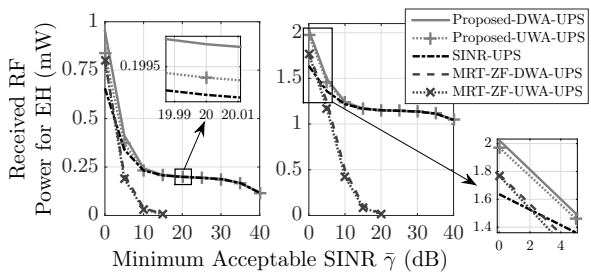} }}
	\vspace{-2mm}\caption{\small Received RF power for EH in mW as a function of the SINR $\bar{\gamma}$ in dB for $N=K=4$, $L=5$m, and for the different low complexity sub-optimal schemes.}
	\label{fig:comparison} 
\end{figure}

In Fig.~\ref{fig:comparison} we plot the received RF power for EH in mW versus $\bar{\gamma}$ in dB for $N=K=4$, $L=5$m, and for all 
under comparison designs. It is evident that both our proposed low complexity designs and SINR-UPS significantly outperform MRT-ZF-UWA-UPS and MRT-ZF-DWA-UPS. The gap averaged over all SINR demands between the EH power achieved by DWA and UWA is less than $-17$dBm, in other words, the average improvement of DWA over UWA is around $2\%$. This gap between DWA and SINR-UPS is around $-13$dBm, hence, the corresponding average improvement is around $7\%$.
\begin{figure}[!t]
	\centering 
	{{\includegraphics[width=3.1in]{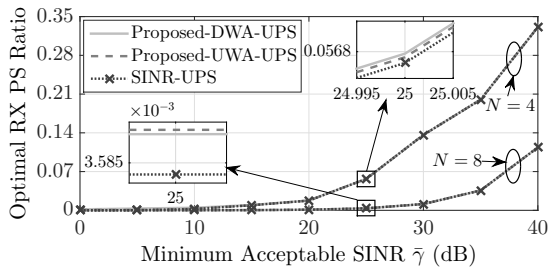} }} 
	\vspace{-2mm}\caption{\small Optimal UPS $\bar{\rho}^*$ ratio as a function of the SINR $\bar{\gamma}$ in dB for the proposed designs with DWA and UWA, as well as SINR-UPS~\cite{MU-MISO-TBF-PS} with $K=4$, $L=5$m, and different $N$ values.}
	\label{fig:PS-ratio2} 
\end{figure} 
In Fig.~\ref{fig:PS-ratio2} we plot the optimal UPS ratio $\bar{\rho}^*$ versus $\bar{\gamma}$ in dB for our two proposed designs and SINR-UPS considering $K=4$, $L=5$m, and different $N$ values. It is obvious that $\bar{\rho}^*$ is very similar for all three designs, a fact that justifies their similar achieved EH power in Fig.~\ref{fig:comparison}.

\begin{figure}[!t]
	\centering 
	{{\includegraphics[width=3.1in]{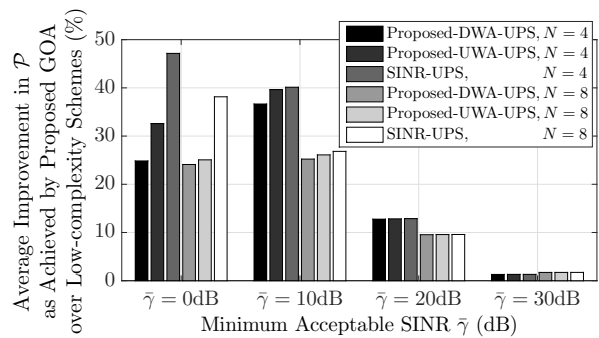} }}
	\vspace{-2mm}\caption{\small Percentage of the average performance improvement in received RF power $\mathcal{P}$ for EH as a function of the SINR $\bar{\gamma}$ in dB of GOA over the proposed low complexity sub-optimal designs for $K=4$, $L=5$m, and different $\bar{\gamma}$ and $N$ values. The plotted values represent the improvement of GOA over  other low complexity schemes with highest improvement achieved over SINR-UPS scheme.}
	\label{fig:enhancement} 
\end{figure}
The performance comparison of our GOA and low complexity sub-optimal designs together with SINR-UPS is included in Fig.~\ref{fig:enhancement}, where $K=4$, $L=5$m, and different values for $\bar{\gamma}$ in dB and $N$ have been considered. As shown, GOA provides for $N=4$ an average improvement of about $19\%,$ $21\%,$ and $26\%$ over the Proposed-DWA-UPS, Proposed-UWA-UPS, and SINR-UPS designs, respectively, in terms of achievable RF power for EH. When the number of TX antennas increases to $N=8$, this performance enhancement slightly reduces to $15\%,$ $15.5\%,$ and $20\%$, respectively. Obviously, despite the relatively high GOA complexity, this algorithm provides sufficient performance improvement for low and medium vales of the SINR demands. However, for high SINR demands, this performance improvement is not as significant. One may also notice that the proposed design adopting DWA that requires $x K$ computations does not provide significant improvement over that based on UWA that requires only $x$ computations. In addition, its is shown in the last two figures that the Proposed-UWA-UPS design outperforms SINR-UPS with an average performance improvement of around $5\%$.
\begin{figure}[!t]
	\centering 
	{{\includegraphics[width=3.1in]{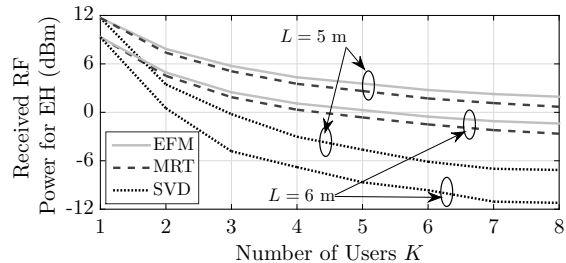} }}
	\vspace{-2mm}\caption{Received RF power for EH in dBm as a function of the RXs' number $K$ for the different low complexity sub-optimal designs and relevant benchmark designs with $N=8$ and different $L$ values.}
	\label{fig:energy}  
\end{figure} 

In Fig$.$~\ref{fig:energy} we finally compare for $N=8$ and different $L$ values the received RF power for EH obtained using our proposed EFM TX precoding design presented in Section~\ref{sec:EHP}, the MRT design of~\cite{EBF2}, and the TX energy beamforming design of~\cite{EBF1} that is based on the Singular Value Decomposition (SVD) of the concatenated channel matrix for all RXs. As observed, MRT performs close to our proposed design exhibiting a mean performance degradation of about $1.2$dBm. The SVD design, however, that targets at maximizing the sum RF power for EH performs very poor in terms of EH fairness performance. We thus conclude that not only our proposed joint TX precoding and IoT PS design provides significant improvements over the existing competitive benchmarks schemes, but even our proposed EFM TX precoding designs yields significant energy savings over relevant ones. 

\section{Conclusions}\label{sec:conclusion} 
In this paper, we investigated the max-min EH fairness problem in MISO SWIPT multicasting IoT systems comprising of PS IoT devices having individual QoS constraints. A generic RF EH model that captures practical rectification operation was adopted. We first obtained an equivalent SDR formulation for the considered design problem and then presented an efficient algorithmic implementation for the jointly globally optimal TX precoding and IoT PS ratio parameters. It was shown that each optimal TX precoding vector has a special regularized ZF structure, based on which a novel weighted TX beamforming direction was proposed for serving each IoT device. Tight closed form approximations for the optimal TX PA allocation and RX UPS ratio were derived for a given weighted TX beamforming direction. Our extensive numerical investigations validated the presented analysis and verified the importance of the proposed design, while showcasing the interplay of critical system parameters. Selected results showed that the proposed jointly optimal design outperforms the existing benchmark ones, while yielding a significant performance gain of more than $20\%$ over the nearest competitor.  {Future extensions of the presented framework include the consideration of multiple antennas at the IoT devices and massive antenna arrays at TX, as well as of millimeter wave applications with hybrid beamforming architectures.}  

\bibliographystyle{IEEEtran} 
\bibliography{refs_MU_MISO_PS_TBF}
\end{document}